\documentclass{aa}  
\usepackage{amssymb}
\usepackage{graphicx}
\usepackage{hyperref}
\usepackage{xcolor}
\newcommand*{\bba}{$^{\scriptstyle 3\mathrm{D}}$B{\sc {arolo}}}

\usepackage{txfonts}
\titlerunning{Dynamics from warm and cold gas tracers}

\begin{document}

   \title{The ALMA-ALPAKA survey II. Evolution of turbulence in galaxy disks across cosmic time: Difference between cold and warm gas}


   \author{F. Rizzo
          \inst{1,2,3}
          \and
          C. Bacchini
          \inst{4}
          \and
          M. Kohandel
          \inst{5}
          \and
          L. Di Mascolo
          \inst{6}
          \and
          F. Fraternali
          \inst{3}
          \and
          F. Roman-Oliveira
          \inst{3}
          A. Zanella
          \inst{7}
          \and
          G. Popping
          \inst{8}
          \and
          F. Valentino
          \inst{8}
          \and
          G. Magdis
          \inst{1, 9}
          \and
          K. Whitaker
          \inst{10}
          }

    \institute{Cosmic Dawn Center (DAWN)
   \and Niels Bohr Institute, University of Copenhagen, Jagtvej 128, 2200 Copenhagen N, Denmark
    \and  Kapteyn Astronomical Institute, University of Groningen, Landleven 12, 9747 AD, Groningen, The Netherlands \\
    \email{rizzo@astro.rug.nl}
   \and Dark, Niels Bohr Institute, University of Copenhagen, Jagtvej 128, DK-2200 Copenhagen, Denmark
   \and Scuola Normale Superiore, Piazza dei Cavalieri 7, I-56126 Pisa, Italy
   \and 
   Laboratoire Lagrange, Université Côte d'Azur, Observatoire de la Côte d'Azur, CNRS, Blvd de l'Observatoire, CS 34229, 06304 Nice cedex 4, France
    \and
    Istituto Nazionale di Astrofisica, Vicolo dell'Osservatorio 5, 35122, Padova, Italy
    \and
    European Southern Observatory, Karl-Schwarzschild-Str. 2, D-85748 Garching bei München, Germany
    \and
    DTU-Space, Technical University of Denmark, Elektrovej 327, DK-2800, Kgs. Lyngby, Denmark
    \and
    Department of Astronomy, University of Massachusetts, Amherst, MA 01003, USA    
    }


\abstract{
The gas in the interstellar medium (ISM) of galaxies is supersonically turbulent. Measurements of turbulence typically rely on cold gas emission lines for low-$z$ galaxies and warm ionized gas observations for $z > 0$ galaxies. Studies of warm gas kinematics at $z > 0$ conclude that the turbulence strongly evolves as a function of redshift, due to the increasing impact of gas accretion and mergers in the early Universe. 
However, recent findings suggest potential biases in turbulence measurements derived from ionized gas at high-$z$, impacting our understanding of turbulence origin, ISM physics and disk formation.
We investigate the evolution of turbulence using velocity dispersion ($\sigma$) measurements from cold gas tracers (i.e., CO, [CI], [CII]). The initial dataset comprises 17 galaxy disks with high data quality from the ALPAKA sample, supplemented with galaxies from the literature, resulting in a sample of 57 galaxy disks spanning the redshift range $z = 0 - 5$. This extended sample consists of main-sequence and starburst galaxies with stellar masses $\gtrsim 10^{10} M_{\odot}$. The comparison with current H$\alpha$ kinematic observations and existing models demonstrates that the velocity dispersion inferred from cold gas tracers differ by a factor of $\approx 3$ from those obtained using emission lines tracing the warm, ionized gas. We show that stellar feedback is the main driver of turbulence measured from cold gas tracers and the physics of turbulence driving does not appear to evolve with time. This is fundamentally different from the conclusions of studies based on warm gas, which had to consider additional turbulence drivers to explain the high values of $\sigma$. 
We present a model predicting the redshift evolution of turbulence in galaxy disks, attributing the increase of $\sigma$ with redshift to the higher energy injected by supernovae due to the elevated star-formation rate in high-$z$ galaxies. This supernova-driven model suggests that turbulence is lower in galaxies with lower stellar mass compared to those with higher stellar mass. Additionally, it forecasts the evolution of $\sigma$ in Milky-Way like progenitors.
}


   \maketitle
%

\section{Introduction} \label{sec:intro}
The gas within the interstellar medium (ISM) of both low- \citep[e.g.,][]{Elmegreen_2004, Utomo_2019, Bacchini_2020} and high--$z$ \citep[e.g., see][for a review]{Forster_2020} galaxies is known to be supersonically turbulent. 
The strongest observational evidence for the presence of turbulent motions within the ISM is that the gas velocity dispersion $\sigma$, which measures the broadening of emission lines, exceeds the expected broadening from thermal motions. This additional broadening allows for quantifying the random, chaotic motions 
of the gas in the ISM. 

For $z \approx 0$ galaxies, turbulence is usually measured through spatially-resolved observations of emission lines tracing the cold molecular and atomic gas \citep[e.g., CO, HI][]{Ianjamasimanana_2015, Mogotsi_2016, Utomo_2019, Bacchini_2020}. For $z > 0$ galaxies, obtaining spatially-resolved observations of cold gas emission lines requires exceptionally long integration times \citep{Rizzo_2023}, even with the Atacama Large Millimiter Array (ALMA). Consequently, studies of gas turbulence at $z > 0$ have mainly relied on Integral Field Unit (IFU) or long-slit observations of emission lines tracing warm, ionized gas \citep[e.g., H$\alpha$, {[}OII{]}, {[}OIII{]};][]{Epinat_2009, Stott_2016, Wisnioski_2015, DiTeodoro_2016, Simons_2016, Turner_2017}, that were possible only up to $z\approx3$ before the launch of JWST. Crucially, these studies based on ionized gas measurements have assumed that, for high-$z$ galaxies, the ionized gas kinematics (i.e., $\sigma$ and rotation velocity $V$) is consistent with the ones derived from cold gas \citep[e.g.,][]{Genzel_2011, Wisnioski_2015, Johnson_2018}. 

One of the main results from warm gas kinematic studies is that the turbulence strongly evolves as a function of redshift \citep[e.g.,][]{Epinat_2009, Forster_2009, Wisnioski_2015, Simons_2017, Turner_2017, Johnson_2018, Birkin_2023, Puglisi_2023}. Galaxy disks at $z = 2 - 3$ are more turbulent and less rotationally supported (i.e., values of rotation-to-velocity dispersion ratio, $V/\sigma \lesssim 5$) than galaxy disks at $z < 1$. These results have been recently challenged by an increasing number of kinematic studies obtained from [CII] observations of $z \gtrsim 4$ galaxies \citep[e.g.,][]{Neeleman_2020, Rizzo_2020, Rizzo_2021, Lelli_2021, Tsukui_2021, Pope_2023, Posses_2023, Roman_2023} using ALMA. These works find that the gas turbulence within $z > 4$ galaxy disks is, on average, lower than those expected by extrapolating the evolution of the $\sigma - z$ relation obtained from warm gas studies at $z = 0 - 3$ \citep{Ubler_2019} and that $z \sim 4$ disks are dynamically cold, with $V/\sigma$ values of $\approx 10$ \citep[e.g.,][]{Neeleman_2020, Rizzo_2020, Fraternali_2021, Rizzo_2021, Lelli_2021, Tsukui_2021, Pope_2023, Roman_2023}. 

The above discrepancy between the kinematics obtained from [CII] observations and those extrapolated using ionized gas could be due to intrinsic differences between the cold and warm gas components. At $z = 0$, the values of $\sigma$ from cold gas (e.g., CO, HI) are, in fact, systematically lower than the ones obtained from H$\alpha$ \citep[e.g.,][]{Epinat_2008, Varidel_2020, Law_2022}. Moreover, 
\citet{Kohandel_2024} use synthetic observations of zoom-in simulations of galaxies at $z > 4$ to highlight a potential additional bias in turbulence measurements derived from warm gas velocity dispersion. In their studies, unresolved non-circular motions induced by outflows contaminate the measurements of turbulence from warm gas. As a result, warm gas $\sigma$ values are $2-3$ times higher than the cold gas ones. In other words, in addition to the thermal and turbulent motions, the shape of the emission lines tracing the warm gas can be affected by the presence of non-circular (radial or vertical) motions, induced by outflows. These motions can mimic the broadening due to turbulence when the spectral or angular resolutions of the data are suboptimal.

In addition to the different kinematic tracers, potential contamination from outflows, and the redshift range (i.e., $1 - 3$ for warm gas and $> 4$ for cold gas), there are two additional potential reasons that can explain the discrepancy between warm and cold gas kinematic studies at $z > 0$: intrinsic differences in the galaxy populations or biases due to observational limitations. First, the results obtained with warm gas rely on observations of main-sequence galaxies \citep{Rodighiero_2011, Whitaker_2012} with a wide range of stellar masses ($M_{\star} = 10^{9} - 10^{10} M_{\odot}$), while the [CII] studies at $z > 4$ have focused on massive ($M_{\star} \gtrsim 10^{10} M_{\odot}$) upper main-sequence or starburst galaxies. Second, the typical angular resolution of ground-based telescopes used to observe $z = 1 - 3$ galaxies
\citep[e.g., $\gtrsim 0.5$'' in seeing-limited mode][]{Wisnioski_2015, Harrison_2017, Turner_2017, Puglisi_2023} poses challenges due to beam smearing, potentially leading to degeneracies between rotation velocity and velocity dispersion \citep{DiTeodoro_2015, Rizzo_2022}. Additionally, even when the galaxies are well spatially resolved thanks to JWST, the spectral resolutions of current facilities (e.g., FWHM $\gtrsim$ 100 km/s for JWST/NIRSpec) complicate the kinematic analysis, hampering the possibility of both cleanly disentangling between turbulent and non-circular motions and measuring values of $\sigma \lesssim$ FHWM/2.35 $\approx$ 40 km/s. On the other hand, cold gas emission lines usually fall in the sub-mm to radio range of the electromagnetic spectrum. In these wavelength ranges, the typical spectral resolutions of instruments is of a few km/s for low-$z$ data and 30 km/s for high-$z$ data. Combined with the typical angular resolution of $\approx$0.2", this kind of observations allows one to identify or isolate the circular motions in the rare cases when cold gas outflows are present \citep{Rizzo_2023, Roman_2023}.

Understanding whether the origin of the discrepancy between studies based on warm and cold gas tracers is driven by different galaxy 
 populations or biases due the kinematic tracers and observational limitations has significant implications on galaxy formation and evolution, even besides the theories of disk formation. For instance, the clumpy morphology of high-$z$ galaxies and the formation of galaxy bulges at high-$z$ have been explained within the framework of highly turbulent disks \citep{Bournaud_2010}. In this scenario, mergers and vigorous gas accretion
give rise to highly turbulent motions and instabilities within disks. These instabilities cause fragmentation of the gas disk, the formation of massive star-forming regions, and their migration into the inner regions \citep{Dekel_2013, Bournaud_2014}. Finally, inaccurate measurements of the velocity dispersion can significantly affect the derivation 
of the mass and distribution of dark matter in galaxies, as their rotation curves must be corrected for the asymmetric drift which depends on $\sigma$ \citep[see][for details]{Lelli_2022}.

Given the aforementioned potential biases on the $\sigma$ measured using the warm, ionized gas, key questions arise: Do the cold gas kinematics exhibit differences compared to warm counterparts when galaxies with similar physical properties are considered? Do the dynamical properties of galaxies depend on the gas tracers used to probe the gas kinematics?
 Does turbulence in disks evolve with cosmic time when measured from cold gas tracers?
    
In this paper, we address these questions 
by analyzing dynamical properties obtained from cold gas tracers across a redshift range from 0 to 5. The paper is structured as follows: in Section~\ref{sec:sample}, we describe the sample used in our analysis; in Section \ref{sec:evolution}, we provide an empirical parametrization for the redshift evolution of $\sigma$ and $V/\sigma$ and we compare it with ionized gas studies from the literature. In 
 Section \ref{sec:drivers}, we briefly discuss the potential drivers of turbulence, while in Section \ref{sec:model}, we present a physically-motivated model, calibrated on observations, predicting the turbulence evolution in star-forming galaxies. Caveats regarding the available values of $\sigma$ and star-formation rate (SFR) and a summary of this work are discussed in Sections \ref{sec:caveats} and \ref{sec:conclusions}, respectively.

\section{ALPAKA and the extended sample} \label{sec:sample}
The primary objective of this paper is to study turbulence through the kinematics of cold gas across the widest possible redshift range. 
While previous works have explored turbulent motions and their driving mechanisms at $z > 0$, the focus has largely been on the range of $z = 4 - 5$ \citep[e.g.,][]{Neeleman_2020, Rizzo_2021, Tsukui_2021, Roman_2023}. This particular redshift interval offers favorable conditions for conducting systematic, high-resolution observations crucial for accurate kinematic measurements, primarily due to the brightness of the [CII]-158$\mu$m emission line \citep{Carilli_2013}. Observational campaigns targeting the intermediate-redshift range (i.e., $z = 0.5 - 4$) are significantly challenging. In fact, observations of [CII] are almost unfeasible because of the limited atmospheric transmission in the frequency range of the redshifted [CII] line and the restricted coverage of ALMA bands \citep{Carilli_2013}. CO or [CI] emission lines are promising alternatives for probing the kinematics of cold gas within this intermediate redshift range. Nonetheless, the faintness of CO and [CI] emission lines poses significant observational challenges. Even with ALMA's capabilities, long integration time (i.e., $\gtrsim 20$ hours) are typically needed to spatially resolve the CO emission in normal main-sequence galaxies at $z \sim 2$. Consequently, despite a decade of ALMA operations, there have been no high-resolution systematic surveys of CO at $z = 0.5 - 4$, but only observations of single targets.

In response to this limitation, we initiated the "ALMA Archival Large Program to Advance Kinematic Analysis" (ALMA-ALPAKA) project \citep{Rizzo_2023}. ALPAKA is based on the mining of the ALMA archive to gather high-quality observations of CO or [CI] emission lines from non-lensed, star-forming galaxies within the redshift range of $z = 0.5 - 4$. The ALPAKA project encompasses $\approx$0.25" observations for 28 massive star-forming galaxies, representing the most extensive $z = 0.5 - 4$ sample, to date, with spatially-resolved cold gas kinematics.

In \citet{Rizzo_2023}, we used the ALMA data to infer the dynamical state of the ALPAKA galaxies, and then divide the sample in three kinematic classes: rotating disks (19/28), interacting systems (2/28), uncertain (7/28). Rotation velocity ($V$) and velocity dispersion ($\sigma$) profiles derived using \bba\, \citep{DiTeodoro_2015} are presented and discussed in detail in \citet{Rizzo_2023}. 

For the analysis presented in this paper, we selected the 17 ALPAKA disks with robust (i.e., uncertainties $\lesssim 60\%$) estimates of their stellar masses $M_{\star}$ and SFR \citep{Rizzo_2023}. Due to the aforementioned observational limitations, the 17 ALPAKA disks are representative of the star-forming galaxy population, with stellar masses $M_{\star} \gtrsim 10^{10} M_{\odot}$. Most of the galaxies that are included in the ALPAKA sample were discovered as bright sources in the infrared or submillimeter wavelength \citep[see][for details]{Rizzo_2023}. The 17 ALPAKA galaxies have SFRs that are 1.3 to 28 times higher than the SFR of a main-sequence galaxy (SFR$_{\mathrm{MS}}$).  We quantify this offset  $\Delta_{\mathrm{MS}}= \mathrm{SFR}/\mathrm{SFR_{MS}}(M_{\star}, z)$ using the prescriptions recently published by \citet{Popesso_2023} and normalized to a Chabrier Initial Mass Function \citep[IMF,][]{Chabrier_2003}:

\begin{align} 
    MS(M_{\star}, z) \equiv \log \left[\frac{\mathrm{SFR_{MS}}(M_{\star}, z)}{M_{\odot} yr^{-1}} \right] = 
    \begin{aligned}[t]
    &(-0.034\,t(\mathrm{Gyr}) + 4.722) \times m \\
    &-0.1925 \times m^2 + \\
    & + (-26.16+0.2\,t(\mathrm{Gyr})),
    \end{aligned}
    \label{eq:sfr}
\end{align}
where $t$ is the cosmic time and 
\begin{equation*}
m = \left[\log \left( \frac{M_{\star}}{M_{\odot}} \right) + 0.025\right].
\end{equation*}

\begin{figure}[th!]
    \begin{center}  \includegraphics[width=\columnwidth]{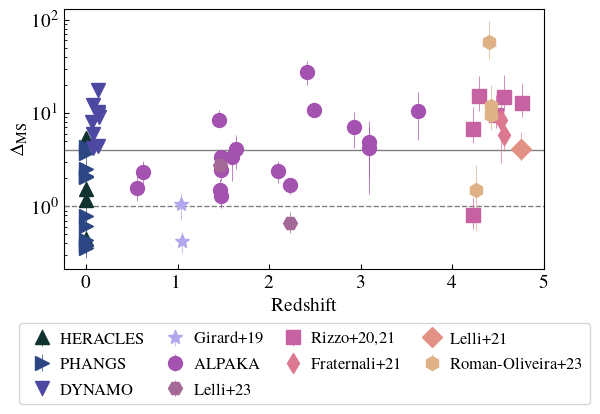}
        \caption{Location of the extended sample in the plane of main-sequence offset $\Delta_{\mathrm{MS}}$ vs redshift. The gray dotted line shows the location of main-sequence galaxies with $\Delta_{\mathrm{MS}} = 1$ for reference \citep{Popesso_2023}. The gray solid line shows the separation between main-sequence and starburst populations at $\Delta_{\mathrm{MS}} = 4$.}  
        \label{fig:ssfr}
    \end{center}
\end{figure}

To trace the evolution of the dynamical properties of galaxies across cosmic time, we supplement the ALPAKA dataset with kinematic measurements obtained from CO, [CI], or [CII] spatially resolved observations from the literature. Throughout the paper, we refer to this sample as the extended sample. This includes only galaxies that are classified as rotating disks \citep[see the defintion in][]{Rizzo_2023} and with stellar masses $M_{\star} \gtrsim 10^{10} M_{\odot}$ to ensure consistency with the stellar mass range covered by the ALPAKA sample. In Table \ref{tab:tab1}, we summarize the properties of the extended sample, which is described in details in Sects.~\ref{sec:lowz} and \ref{sec:highz}. 
Figure \ref{fig:ssfr} displays the distribution of $\Delta_{\mathrm{MS}}$ versus redshift for all the galaxies in the extended sample. These galaxies span both the main-sequence and starburst regimes \citep[i.e., $\Delta_{\mathrm{MS}} > 4$,][]{Rodighiero_2011}. 
In Table \ref{tab:deltams}, we list the 50th, 2nd and 98th percentiles of the $\Delta_{\mathrm{MS}}$ distributions for the galaxies included in the extended sample (see Appendix \ref{sec:details_sample} for further details). This summary statistics allow us to quantify the spread of the $\Delta_{\mathrm{MS}}$ values in three redshift bins: $0 - 0.13$, $0.5 - 3.5$, $4 - 5$. While the three bins contain both main-sequence and starburst galaxies, the redshift interval $z = 3 - 5$ is skewed towards the starburst region. This bias stems from a selection effect, as at $z > 3$ high-resolution observations are available mostly for bright dusty galaxies. The  $\sigma$ used in this paper refer to the values obtained as a radial average of the velocity dispersion profiles for each galaxy in the extended sample (see Appendix~\ref{sec:details_sample}). The $V/\sigma$ values are obtained as a ratio between the maximum value of the rotation velocity and the average velocity dispersion. 

\begin{table*}[h!] 
\begin{center}
\caption{Datasets included in the extended sample.
}
\label{tab:tab1}
\begin{tabular}{ccccc}
\\
\hline\hline \noalign{\smallskip}
Name  &  Number of galaxies & Redshift & Kinematic tracer & Reference for the kinematic analysis  \\
\noalign{\smallskip}
\hline
\noalign{\medskip}
HERACLES & 4 & 0 & CO(2-1) & \citet{Bacchini_2020}\\
PHANGS & 10 & 0 & CO(2-1) & This work\\
DYNAMO & 9 & 0.07 - 0.13 & CO(3-2), CO(4-3) & \citet{Girard_2021} \medskip \\
ALPAKA & 17 & 0.5 - 3.5 & CO and [CI] lines & \citet{Rizzo_2023}\\
Girard+19 & 2 & 1.03, 1.04 & CO(4-3) & \citet{Girard_2019}\\
Lelli+23 & 2 & 1.46, 2.23 & CO(4-3) & \citet{Lelli_2023} \medskip \\
Rizzo+20, 21 & 6 &  4 - 5 & [CII] & \citet{Rizzo_2021}\\
Fraternali+21 & 2 & 4.54, 4.57 & [CII] & \citet{Fraternali_2021}\\
Lelli+21 & 1 & 4.76  & [CII] & \citet{Lelli_2021}\\
Roman-Oliveira+23 & 4 & 4 - 5  & [CII] & \citet{Roman_2023}\\

\noalign{\smallskip}
\hline
\end{tabular}
\end{center}
\end{table*}

\begin{table}[h!]
    \centering
    \caption{Percentiles of the $\Delta_{\mathrm{MS}}$ distributions for the galaxies in the extended sample, divided in three redshift bins.}
    \begin{tabular}{c|cccc}
        \hline
        \hline
        Subsample & Redshift range & 50th & 2nd & 98th
        \\
        \hline
        Low-$z$ & $0.5 - 3.5$ & 3.6 & 0.4 & 15\\
        Intermediate-$z$ & $0.5 - 3.5$ & 2.8 & 0.5 & 20\\
        High-$z$ & $4 - 5$ & 9 & 1 & 46\\
        \noalign{\smallskip}
        \hline       
    \end{tabular}   
    \label{tab:deltams}
\end{table}

\subsection{Low-z subsample}\label{sec:lowz}
In the low-$z$ subsamples, we select 23 galaxies with kinematic data obtained from CO observations and $\Delta_{\mathrm{MS}} \gtrsim 0.3$, matching their properties with those of galaxies at $z \gtrsim 0.5$. Throughout the paper, we will denote the galaxies in this subsample by the name of the surveys to which they belong, as outlined below.

HERACLES: \citet{Bacchini_2020} analyzed the CO(2-1) kinematics of 7 disk galaxies from the  Heterodyne Receiver Array CO Line Extragalactic Survey \citep[HERACLES,][]{Leroy_2009} using \bba\ \citep{DiTeodoro_2015}. In our extended sample, we include the 4 galaxies from \citet{Bacchini_2020} with stellar masses $M_{\star} \gtrsim 10^{10} M_{\odot}$. 

PHANGS: We include ten galaxies from the Physics at High Angular resolution in Nearby Galaxies (PHANGS)-ALMA survey \citep{Leroy_2021}. PHANGS-ALMA maps the CO(2-1) emission 
from a sample of nearby star-forming galaxies. The velocity $V$ and velocity dispersion $\sigma$ values used in this paper are derived after applying \bba\ \citep{DiTeodoro_2015} to the publicly available data cubes \citep[][see Appendix \ref{sec:details_sample} for details]{Leroy_2021, Leroy_2021b}.

DYNAMO: The DYnamics of Newly Assembled Massive Objects (DYNAMO) survey \citep{Green_2014} includes massive ($M_{\star} > 10^{10} M_{\odot}$) disk galaxies at $z \approx 0.1$ with properties akin to main-sequence galaxies at $z = 1.5 - 2$. DYNAMO galaxies are starbursts, exhibiting high SFR ($\approx$ 10 M$_{\odot}$/yr) and high gas fractions \citep[20-40 \%][]{Fisher_2014, Fisher_2017}. In our sample, we include the 9 DYNAMO galaxies with gas kinematics derived from CO(3-2) and CO(4-3) emission lines \citep{Girard_2021}.

\subsection{High-z subsamples}\label{sec:highz}
For galaxies at $z \gtrsim 0.5$, in addition to the stellar mass-based selection criterion, we imposed a criterion based on the quality of observational data. We specifically selected galaxies from the literature covered by at least 3 independent resolution elements along the major axis of the gas disk. This ensures the reliability of $V$ and $\sigma$ measurements and allows us to minimize contamination from mergers and beam smearing \citep{Rizzo_2022}. This quality criterion is not needed for the selection of the low-$z$ samples presented in Sect.~\ref{sec:lowz} as their major axes are typically resolved with more than ten independent resolution elements. 

In addition to the ALPAKA galaxies, the sample at $ 0.5 < z < 4$ comprises two gravitationally lensed galaxies at $z \approx 1$, whose kinematic analysis from CO(4-3) is detailed in \citet{Girard_2019}, and two main-sequence galaxies at $z = 1.46$ and $2.23 $, with CO(4-3) kinematics presented by \citet{Lelli_2023}.

At $z > 4$, we identified 6 lensed \citep{Rizzo_2020, Rizzo_2021} and 7 non-lensed disk galaxies \citep{Fraternali_2021, Lelli_2021, Roman_2023} with spatially-resolved [CII] observations. To the best of our knowledge, there is currently no availability of resolved (i.e., at least 3 resolution elements along the major axis) [CII] observations for disk galaxies at $z \gtrsim 5$.

\subsection{Considerations on the extended sample}\label{sec:3}
The studies of the dynamical properties of galaxies and the impact of different astrophysical mechanisms (e.g., environment, stellar mass, SFR) on the ISM properties requires the analysis of a sample spanning a large range of physical properties. However, the current limitations in sample size and statistics hinder such investigations at present.  
Only with the next generation of facilities (e.g., ALMA upgrade, ngVLA), these studies can be conducted with greater precision \citep{Carpenter_2020, Kadler_2023}, allowing for a more comprehensive understanding of the evolution of galaxy dynamics across cosmic time. Nevertheless, leveraging the sample compiled in this work offers a promising avenue to initiate these studies. Despite comprising galaxies selected through various methods (e.g., sub-mm, infrared, or optical observations), the extended sample covers a narrow range in stellar mass ($M_{\star}\approx10^{10} - 10^{11} M_{\odot}$) but 4 orders of magnitude in SFR.
These characteristics allow us to investigate the evolution of the dynamics of the massive galaxy population. Furthermore, the extended sample allows us to minimize observational biases (e.g., beam-smearing effect, inclusion of non-resolved interacting systems) as it includes only observations with high data quality.


  \section{Redshift evolution of galaxy dynamics using cold gas tracers} \label{sec:evolution}
  In this section, we use the extended sample to study how $\sigma$ and $V/\sigma$ evolve as a function of redshift (Sect.~\,\ref{sec:bestfit}). We then compare the dynamical properties of our extended sample with previous results from the literature (Sect.~\,\ref{sec:comparison}). 

\subsection{Empirical relations using the extended sample} \label{sec:bestfit}
\begin{figure*}[th!]
    \begin{center}  \includegraphics[width=\textwidth]{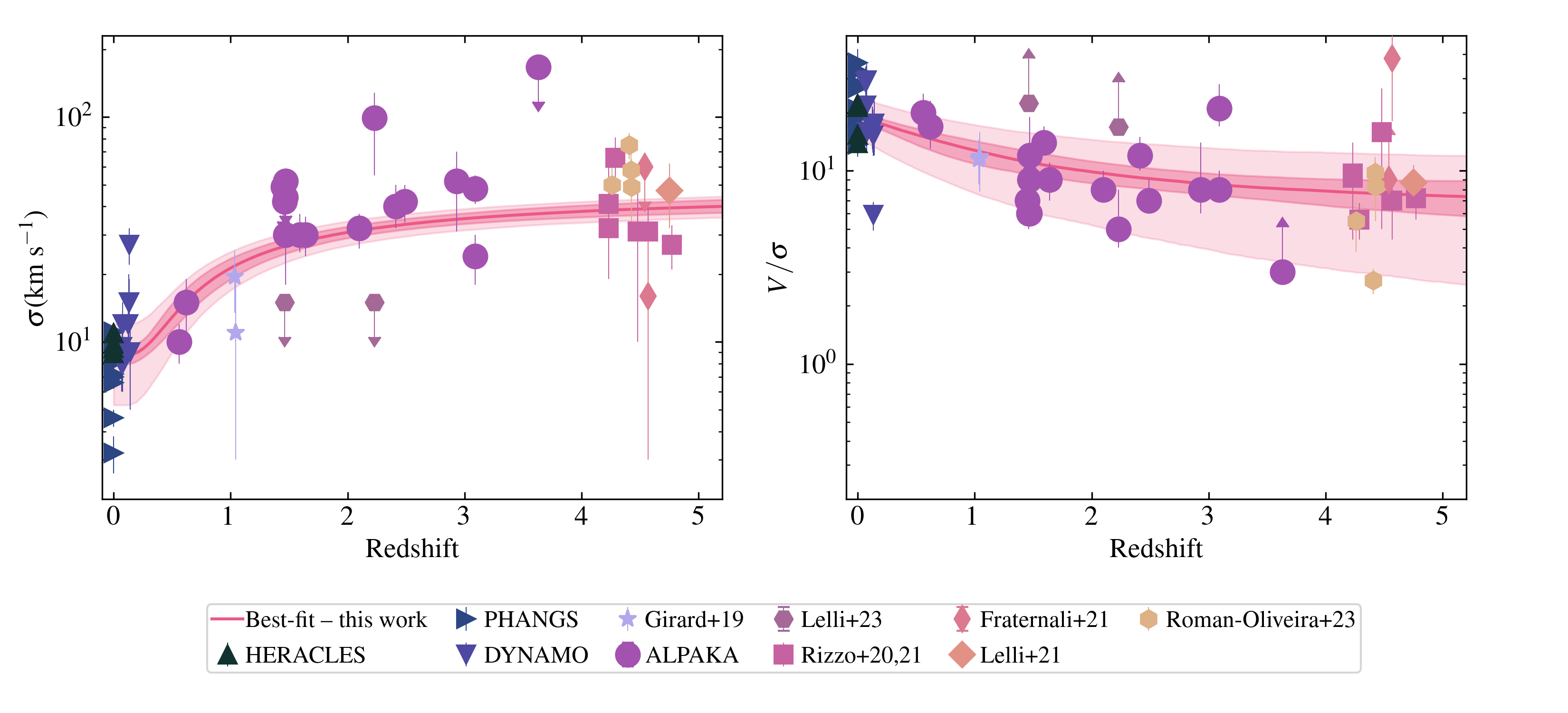}
        \caption{Redshift distribution of velocity dispersion ($\sigma$, left panels) and rotational support ($V/\sigma$, right panel) for our extended sample (colored markers) consisting of massive galaxies. The $\sigma$ and $V/\sigma$ values are derived using emission lines tracing cold gas (i.e., CO, [CI], [CII]). The pink solid lines show the best-fit empirical relations, with the 1-$\sigma$ uncertainties (dark pink area) and the intrinsic scatter (pink area). }  
        \label{fig:vs_cold}
    \end{center}
\end{figure*}

In Fig.~\ref{fig:vs_cold}, we present the distribution of $\sigma$ (left panels) and $V/\sigma$ (right panel) with respect to redshift for our extended sample, represented by colored markers. Despite the scatter, discernible trends emerge in both panels. Specifically, we observe an increase (decrease) by a factor of $\approx 2 - 3$ in $\sigma$ ($V/\sigma$) values in the redshift range $z = 0 - 1$ . Interestingly, these trends become less pronounced at $z > 1$, where a pleatau in both $\sigma$ and $V/\sigma$ is visible.

To quantify these observed trends, we employ the following functional forms:
\begin{equation} \label{eq:sigma}
 \sigma = a_{\sigma} + b_{\sigma} \exp({-z_{0, y}/z}) + s_{\sigma},
\end{equation}
\begin{equation} \label{eq:vsigma}
 V/\sigma = a_{V/\sigma} + b_{V/\sigma} \exp({-z/z_{0, y}}) + s_{V/\sigma}.
\end{equation}
In both equations, the parameters $s_{\sigma}$ and $s_{V/\sigma}$ represent the intrinsic scatter around the median scaling relation. The fits are performed adopting a hierarchical Bayesian method and sampling the parameter space by means of an Hamiltonian Monte Carlo algorithm (see Appendix \ref{sec:fit} for details). Table~\ref{tab:bestfit} reports the best-fit parameters, while the pink solid lines in Fig.~\ref{fig:vs_cold} show the best fit relation with the two dark and shaded regions representing the scatters due to the uncertainties on the parameters and the intrinsic scatters parametrized by $s_{\sigma}$ and $s_{V/\sigma}$. The values of $\sigma$ at $z = 0$, 2, and 4 derived from the best-fit relations are 9, 30, 37 km/s, while the corresponding values of $V/\sigma$ are 18, 9, and 8. The high values of $V/\sigma$ up to $z \sim 5$ indicate that massive galaxy disks are dynamically cold across cosmic time, despite having turbulent motions higher than their local counterparts. 

Our results corroborate and strengthen the conclusions of previous works based on samples covering smaller redshift ranges \citep[e.g.,][]{Neeleman_2020, Rizzo_2020, Fraternali_2021, Lelli_2021, Tsukui_2021, Roman_2023, Lelli_2023}: star-forming massive disk at $z = 0 - 5$ are well settled and rotation-dominated, rather than being highly perturbed by mergers or intense gas accretion.  
It is noteworthy that despite these results being obtained from a sample comprising both main-sequence and starburst galaxies, this conclusion remains conservative, as turbulent motions would be even lower if only main-sequence galaxies were considered (see Sect.~\ref{sec:model} for details).

The discovery of dynamically cold disks at $\sim 4 - 5$ \citep{Neeleman_2020, Rizzo_2020, Rizzo_2021, Lelli_2021} prompted the exploitation of zoom-in simulations \citep{Pallottini_2022} to identify galaxies with similar dynamical properties \citep{Kohandel_2020, Kret_2022, Kohandel_2024}. For instance, \citet{Kret_2022} observed that these dynamically cold disks can form as transient phenomena in their simulations following the accretion of co-planar, co-rotating gas via cold cosmic-web streams. Similarly, \citet{Kohandel_2024} identified dynamically cold disks among their galaxies at $z > 4$. However, owing to the intrinsic characteristics of zoom-in simulations (e.g., limited volumes, resolutions, implementation of feedback presciptions) and the absence of galaxies with stellar masses and SFR comparable to those in the extended sample, it remains challenging to quantify whether state-of-the-art cosmological simulations can produce dynamically cold massive disks with properties (e.g., SFR, $V/\sigma$) similar to those in our extended sample.

\begin{table}[] 
    \centering
    \caption{Best-fit parameters defining the relations between $\sigma$-$z$, Eq~(\ref{eq:sigma}) and $V/\sigma$-$z$, Eq~(\ref{eq:vsigma}).}
    \begin{tabular}{c|cc} 
    \hline
    \hline
    Parameter & $y=\sigma$ & $y=V/\sigma$\\
    \hline
       $a_{y}$  &     9$\pm$ 1 km/s & 7$\pm$2\\
       $b_{y}$  &     39$\pm$5 km/s & 13$\pm$2\\
       $z_{0, y}$ &  1.1$\pm$0.3 & 1.5$\pm$0.9\\ 
       $s_{y}$ &      3.3$\pm$0.9 km/s & 4.5$\pm$0.6\\
       \hline
    \end{tabular}
    \label{tab:bestfit}
\end{table}

  \subsection{Comparison with warm gas kinematics} \label{sec:comparison}

\begin{figure*}[th!]
    \begin{center}  \includegraphics[width=\textwidth]{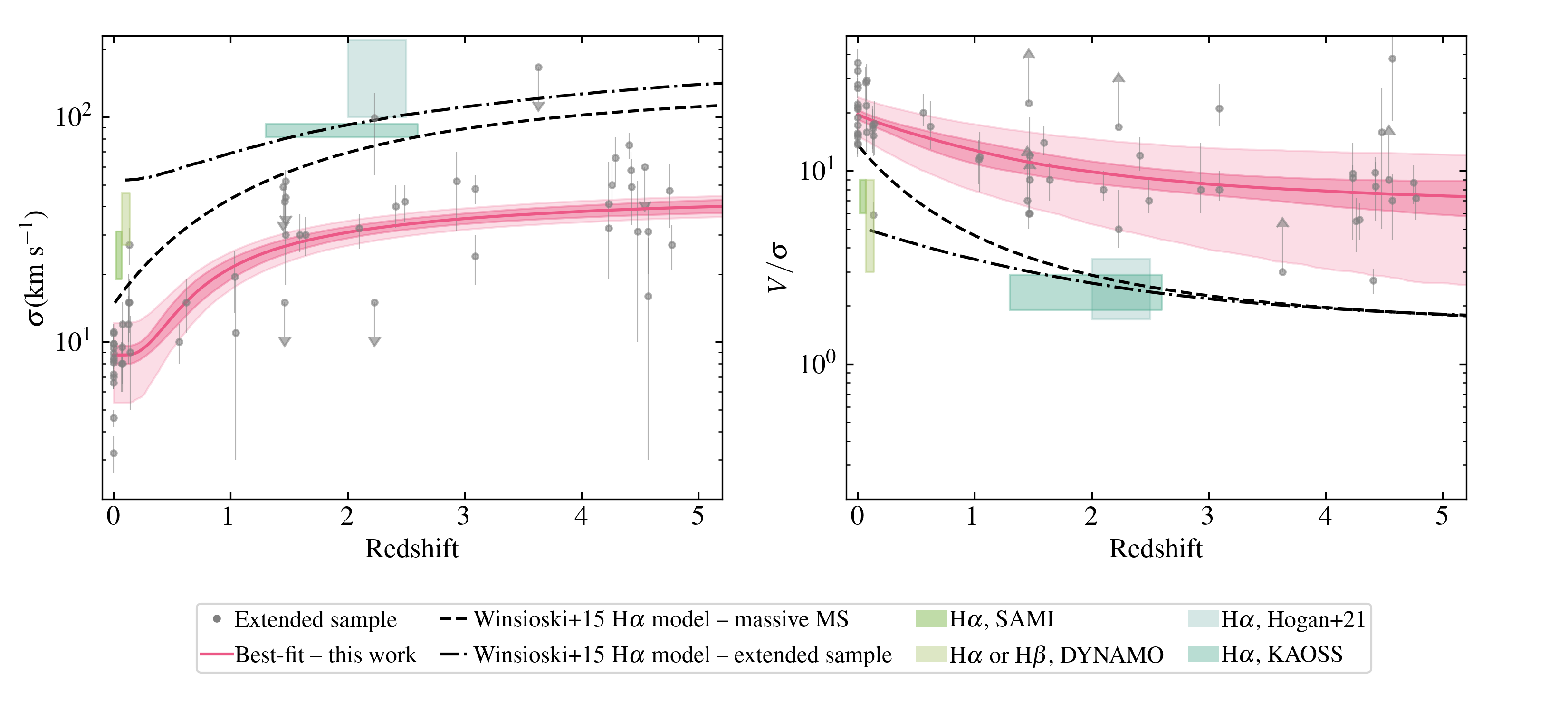}
        \caption{Comparison between cold and warm gas dynamical evolution. The gray points and pink lines show our extended sample and the best-fit relations presented in Fig.~\ref{fig:vs_cold}. The two green boxes indicate the redshift coverages and the 16th and 84th percentiles for samples with galaxy dynamics derived from H$\alpha$ \citep{Varidel_2020, Girard_2021, Hogan_2021, Birkin_2023}.
        The dashed and dash-dotted black lines show the expected relations for samples of main-sequence massive galaxies and for the galaxies in our extended sample, respectively. These relations are derived using the disk-instability model proposed by \citet{Wisnioski_2015}, which is used to reproduce the redshift evolution of $\sigma$ and $V/\sigma$ derived from warm gas tracers (see Sect.\,\ref{sec:comparison}).}  
        \label{fig:vs}
    \end{center}
\end{figure*}

Ideally, one would like to compare the cold and warm gas kinematics in the same sample of galaxies. However, this comparison is hampered by the lack of warm gas observations (e.g., H$\alpha$) for most galaxies in our extended sample. Therefore, we compare the best-fit relations derived in Sect.~\ref{sec:bestfit} with trends of $\sigma$-$z$ and $V/\sigma$-$z$ obtained from warm, ionized gas observations and models available in the literature.

\subsubsection{H$\alpha$ observations}
The redshift evolution of turbulence and $V/\sigma$ from warm gas tracers has been mainly studied using compilations of main-sequence galaxies with a wide range of stellar masses ($\log(M_{\star}/M_{\odot}) = 9 - 11$, see Appendix \ref{sec:ms1} and \ref{sec:ms2}). However, to investigate the impact on the kinematic tracers on the derived $\sigma$ and $V/\sigma$ and minimize potential biases due to differences in physical properties of galaxies, we make comparisons using samples that match the stellar masses and SFRs of our extended sample. In the central columns of Table \ref{tab:comparison}, we summarize the main properties of the different subsamples. For comparison, we report in the right columns the number and the median $\sigma$ and $V/\sigma$ values of galaxies in our extended samples contained in the same redshift bins of the comparison H$\alpha$ subsamples.

In our comparative analysis at $z \approx 0$, we incorporate two distinct subsets. The first subset consists of the 9 DYNAMO galaxies, which are part of our extended sample and for which H$\alpha$ or H$\beta$ kinematics are presented in \citet{Girard_2021}. In the second subset, we included galaxies from the Sydney-AAO Multiobject Integral field Spectrograph (SAMI) survey \citep{Croom_2012, Scott_2018}. In particular, we selected the 67 SAMI galaxies with $M_{\star} \gtrsim 10^{10} M_{\odot}$ and $\Delta_{\mathrm{MS}} \gtrsim 0.3$. Their H$\alpha$ kinematics is presented by \citet{Varidel_2020}. \citet{Girard_2021} have previously studied the comparison between CO and warm gas $\sigma$ values for the 9 DYNAMO galaxies, revealing that, on average, warm gas $\sigma$ values are a factor of 3$\pm1$ greater than those derived from CO. Likewise, the median CO $\sigma$ value within our extended sample at $z \approx 0$ (i.e., PHANGS and HERACLES) is 8$\pm$2 km/s, that is a factor of 3 lower compared to the median H$\alpha$ $\sigma$ of 25$\pm$6 km/s obtained for the SAMI galaxies. The $V/\sigma$ values from CO are on average a factor of 3 higher with respect to the warm gas $V/\sigma$ values for the 9 DYNAMO galaxies. A similar difference of $\approx 3$ is obtained when we compare the median CO $V/\sigma$ from our extended sample to the median H$\alpha$ $V/\sigma$ for the SAMI galaxies.

In our comparative analysis at $z > 0$, we include the $z \sim 2$ samples studied by \citet{Hogan_2021} and \citet{Birkin_2023}.
To our knowledge, there are no systematic H$\alpha$ studies at $z > 4$. 
The sample analyzed by \citet{Hogan_2021} consists of 6 ultra luminous infrared galaxies (ULIRG) at $z = 2.1 - 2.4$, while the one analyzed by \citet{Birkin_2023} consists of 31 dusty star-forming galaxies (DSFGs) at $z = 1.3 - 2.6$. Galaxies in both studies have high stellar masses ($\log(M_{\star}/M_{\odot}) > 10$) and $\Delta_{\mathrm{MS}}$ matching the properties of the intermediate-$z$ subsample (see the percentile values in Table \ref{tab:comparison}). \citet{Hogan_2021} and \citet{Birkin_2023} report median values of $\sigma$ of $160 \pm 60$ km/s and $87 \pm 6$ km/s for the ULIRG and DSFG samples, respectively. The median $V/\sigma$ values are $2.6 \pm 0.9$ for the ULIRGs and $2.4 \pm 0.5$ for the DSFGs (green shaded area in Fig.~\ref{fig:vs}). We now compare these values from H$\alpha$ with the median $\sigma$ and $V/\sigma$ values of our extended sample in the two redshift bins $z = 2.1 - 2.4$ and $z = 1.3 - 2.6$ (see Table \ref{tab:comparison}). 
The values of $\sigma$ from cold gas are a factor of 2 and 4 lower than those found by \citet{Hogan_2021} and \citet{Birkin_2023}, while the corresponding values of $V/\sigma$ from cold gas are a factor of 3 and 4 lower than their H$\alpha$ values. This result clearly indicates that, when considering galaxies with similar stellar masses and $\Delta_{\mathrm{MS}}$, the values of $\sigma$ and $V/\sigma$ from H$\alpha$ are, on average, systematically higher and lower, respectively, than those obtained using cold gas tracers. Providing further support to the findings in this section, we highlight that the systematic disparity of approximately a factor of 3 between cold and warm gas $\sigma$ aligns with observations from the few $z > 1$ galaxies for which both high-resolution CO (or [CII]) and H$\alpha$ data are available \citep{Girard_2021, Parlanti_2023}. 

Due to the data quality of current H$\alpha$ observations (see details in Sect.~\ref{sec:caveats}), we cannot determine what is causing the discrepancy between warm and cold gas kinematics. As a result, we cannot discriminate whether this difference is mainly due to: (i) differences in the disk thickness of the two gas phases; (ii) outflow contaminating the measurements of turbulence from warm gas; (iii) insufficient angular and/or spectral resolution which are biasing the measurements of warm gas $\sigma$. 

\begin{table*}[]
    \centering
    \caption{Main properties of the comparison and extended sample in four redshift bins.}
    \begin{tabular}{c|ccccccc|ccc}
    \hline
    \hline
        & \multicolumn{7}{c}{Comparison samples (warm gas)} & \multicolumn{3}{|c}{Extended sample (cold gas)}\\
        redshift & Name &  Number & \multicolumn{3}{c}{$\Delta_{\mathrm{MS}}$} & $\sigma$ & $V/\sigma$ & Number & $\sigma$ & $V/\sigma$ \\
        & & & 50th & 2nd & 98th & km/s & & & km/s\\
        \hline
        $\approx 0$ & SAMI\tablefootmark{a} &  67 & 1.7 & 0.3 & 6 & 25$\pm6$ & $7^{+2}_{-1}$ & 15 & 8$\pm2$ & $21^{+8}_{-6}$\\
        0.07 - 0.13 & DYNAMO\tablefootmark{b} & 9 & 9 & 4 & 17 & $36^{+10}_{-9}$ & $5^{+4}_{-2}$ & 9 & $12^{+3}_{-4}$ & $17^{+9}_{-2}$\\
        2.1 - 2.4 & Hogan+21\tablefootmark{c} & 6 & 2.6 & 0.8 & 6.2 & 160$\pm60$ & $2.6\pm0.9$ & 4 & $36^{+34}_{-11}$ & $9^{+5}_{-3}$\\
        1.3 - 2.6 &  KAOSS\tablefootmark{d} & 31 & 2.9 & 0.9 & 8.2 & 87$\pm6$ & 2.4$\pm0.5$ & 13 & $40^{+9}_{-11}$ & 10$\pm4$\\
         \hline
    \end{tabular}   
    \label{tab:comparison}
    \tablefoot{Kinematics from:
    \tablefoottext{a}{\citet{Varidel_2020};} \tablefoottext{b}{\citet{Girard_2021};} \tablefoottext{c}{\citet{Hogan_2021};}
    \tablefoottext{d}{\citet{Birkin_2023}}.
    }
\end{table*}

\subsubsection{Models calibrated on H$\alpha$ observations}
In addition to the comparison with the H$\alpha$ samples, we consider the expected evolution of $\sigma$ and $V/\sigma$ according to the disk-instability model proposed by \citet{Wisnioski_2015}. This model is calibrated to reproduce the redshift trends of $\sigma$ and $V/\sigma$ from warm gas observations \citep[e.g.,][]{Wisnioski_2015, Turner_2017, Johnson_2018} and it is based on two assumptions: cold gas disks are marginally unstable to perturbations; and the cold gas $\sigma$ values are similar to the warm gas ones. The presence of instabilities is commonly quantified through measurements of the Toomre parameter $Q$ \citep{Toomre},
\begin{equation} \label{eq:toomre}
Q = \frac{\kappa \sigma}{\pi G \Sigma_{\mathrm{gas}}}.
\end{equation}
In Eq.~(\ref{eq:toomre}), $\Sigma_{\mathrm{gas}}$ is the surface density of the gas and $\kappa$ is the epicyclic frequency which is a function of the angular velocity $\Omega = V/R$. A disk is unstable if $Q \lesssim 1$, that is, when the self-gravity of the gas overcomes the repelling forces due to gas pressure (i.e., $\sigma$) and rotation support ($\Omega$). In \citet{Wisnioski_2015} model, the evolution of the dynamical properties of the star-forming galaxy population is explained in terms of the redshift increase of gas fraction $f_{\mathrm{gas}}$ within marginally unstable disks. In particular, the evolution of $\sigma$ and $V/\sigma$, defined by approximating Eq.~(\ref{eq:toomre}) \citep{Genzel_2011}, are given by the following equations:
\begin{equation}\label{eq:di1}
    \sigma(z) = \frac{Q}{\sqrt{2}} V f_{\mathrm{gas}}(z)
\end{equation}

\begin{equation}\label{eq:di2}
    \frac{V}{\sigma} = \frac{\sqrt{2}}{f_{\mathrm{gas}}(z) Q},
\end{equation}

Following \citet{Wisnioski_2015}, we use Eqs.~(\ref{eq:di1}) and (\ref{eq:di2}) with $Q = 1$, which is the typical value for marginal instability.
We compute the expected evolution of $\sigma$ and $V/\sigma$ in the two cases: 
\begin{enumerate}
    \item massive main-sequence galaxies. We compute $f_{\mathrm{gas}}$ using the empirical relation in Eq.~(3) in \citep{Wisnioski_2015}, which is a function of redshift and stellar mass. We fix the stellar mass at $10^{10.5} M_{\odot}$ and use $V = 200$ km/s, the average rotation velocity typically measured for massive main-sequence galaxies \citep{Harrison_2017, Forster_2018}. The resulting relations are shown in Fig.~\ref{fig:vs} (bottom panels) by a black dashed line.
    \item galaxies from our extended sample. We compute the expected values of $\sigma$ and $V/\sigma$ using Eqs.(\ref{eq:di1}) and (\ref{eq:di2}) ($\sigma_{\mathrm{exp, DI}}$, $(V/\sigma)_{\mathrm{exp, DI}}$) for the galaxies in our extended sample. The values of $\sigma_{\mathrm{exp, DI}}$ and $(V/\sigma)_{\mathrm{exp, DI}}$ represent, therefore, the values predicted by the disk-instability model for galaxies with $V$ and  $f_{\mathrm{gas}}$ equal to the measured values of our extended sample. We then fitted the resulting distributions using Eqs.~(\ref{eq:sigma}) and (\ref{eq:vsigma}) (see Appendix \ref{sec:comparison_di} for details). 
    The dash-dotted lines in Fig.~\ref{fig:vs} are the best-fit relations to the resulting redshift distributions of $\sigma_{\mathrm{exp, DI}}$ and $(V/\sigma)_{\mathrm{exp, DI}}$.
\end{enumerate}

Both the dashed and dash-dotted curves in Fig.~\ref{fig:vs} systematically overestimate the values of $\sigma$ from cold gas by factors of 2 - 4 at $z \gtrsim 1$, while simultaneously underestimating $V/\sigma$ by the same factors. This discrepancy implies that the disk-instability model fails to accurately reproduce the evolution of the dynamical properties inferred from cold gas tracers. It is important to note that, although the instability-model curves are consistent with the observational results from H$\alpha$ kinematics (green areas), this agreement does not imply that the warm gas disks are marginally unstable. First, the instability-model proposed by \citet{Wisnioski_2015} is valid for the cold gas, under the assumption that cold and warm gas have similar kinematics. As shown in Sect.~\ref{sec:comparison}, this assumption is not valid as, on average, the $\sigma$ values from cold gas appear to be a factor of 3 lower than the warm values. Second, \citet{Bacchini_2024} show that the approximation of the Toomre parameter used by \citet{Wisnioski_2015} to derive Eqs.~(\ref{eq:di1}) and ~(\ref{eq:di2}) significantly underestimates the values of $Q$.

\section{Drivers of turbulence} \label{sec:drivers}
Theoretical models and simulations indicate that turbulence decays rapidly relative to the dynamical timescale of the galaxy \citep[e.g.,][]{Maclow_1998, Kim_2015}. 
Therefore, continuous driving mechanisms are necessary to explain the observed level of turbulence. Stellar feedback is expected to be the primary source of turbulence energy in star-forming galaxies \citep{Maclow_2004, Ostriker_2011, Hayward_2017, Orr_2020}. However, a number of  additional processes (e.g., radial gas flows due to gravitational instabilities \citep{Bournaud_2010, Krumholz_2016, Krumholz_2018, Ginzburg_2022}, gas accretion from the cosmic web \citep{Klessen_2010, Forbes_2023, Jimenez_2023}) are considered when the impact of stellar feedback alone is insufficient to account for the observed turbulence.

\citet{Krumholz_2016} and \citet[][K18, hereafter]{Krumholz_2018} developed analytical predictions for the dependence of the cold gas velocity dispersion on SFR
considering two driving mechanisms of turbulence, stellar feedback and radial gas flows (also called transport model). The transport model predicts values of $\sigma$ which are more than 4 times higher than those predicted by the feedback model at fixed SFR, due to the different scaling between $\sigma$ and SFR in the transport and feedback models. In the transport model, $\sigma \propto \mathrm{SFR}$, while $\sigma \propto \mathrm{SFR}^{1/2}$ for the feedback model with a star formation efficiency free to vary as a function of $\sigma$; no relationship between SFR and $\sigma$ is predicted for the feedback model with fixed star formation efficiency (Appendix \ref{sec:assumptionfeedback}). Comparisons of K18 models with warm gas $\sigma$ of $z = 1 - 3$ galaxies suggest that energy injected by stellar feedback is not sufficient to explain the high values of the observed velocity dispersion. In contrast, the transport model is in good agreement with high-$z$ observations derived from warm gas  \citep[see also][]{Johnson_2018, Ubler_2019}. The feedback model alone reproduces the observations at low-$z$. The fiducial model proposed by K18 is, therefore, a combination of the feedback and transport models and it can reproduce both low and high-$z$ galaxies.

\begin{figure*}[th!]
    \begin{center}  \includegraphics[width=\textwidth]{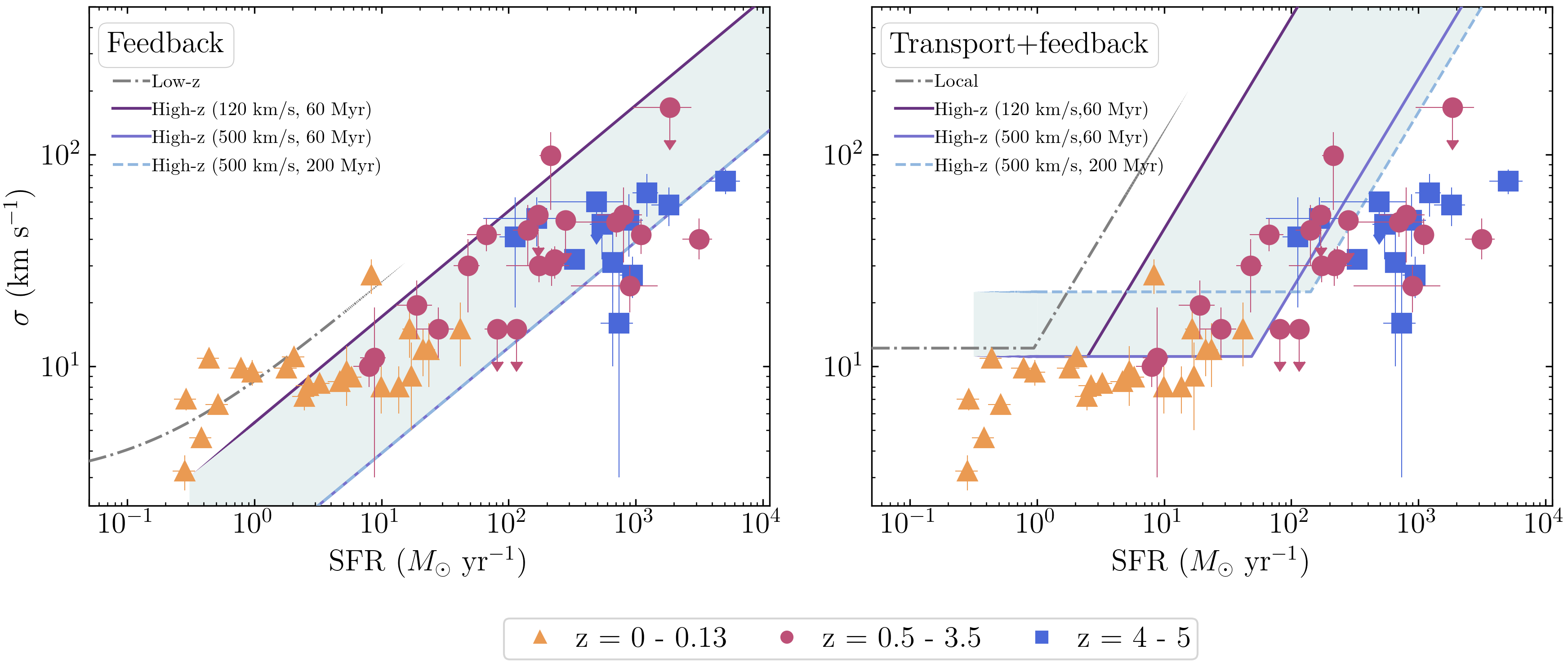}
        \caption{Distribution of $\sigma$ vs SFR for our extended sample divided in three redshift bins, as indicated in the legend, compared with the feedback model (left panel) and the trasport+feedback model (right panel) from K18. The gray dashed and colored solid lines are obtained using the prescriptions for low and high-$z$ galaxies, as reported in K18. For the high-$z$ models, we vary the rotation velocity and orbital times (parenthesis in the legend) in the range covered by the extended sample (shaded area). In the left panel, the solid dark blue and dashed light blue lines overlap. }  
        \label{fig:krum}
    \end{center}
\end{figure*}

In this context, we compare the values of cold gas $\sigma$ with the fiducial and feedback models proposed by K18, in order to understand what is the potential origin of the cold gas turbulence in our extended sample. 
In Fig.~\ref{fig:krum}, we show the distribution of the extended sample in the $\sigma$-SFR plane and the feedback and transport+feedback models. In both models, we assume the fiducial parameters (i.e., rotation velocities and orbital times) for high-$z$ galaxies reported by K18 and we use the range of values of our extended sample (shaded area).

The $\sigma$-SFR distribution of our sample aligns with the predictions of the stellar feedback model proposed by K18. Conversely,
the transport+feedback model predicts, that high-$z$ galaxies with SFR in the range of the extended sample have $\sigma$ values which are, on average, $3-10$ times higher than the values of $\sigma$ of our galaxies. The findings presented in this section corroborate previous studies showing that the turbulence measured from cold gas kinematics can be explained  by stellar feedback \citep{Girard_2021, Rizzo_2021, Roman_2024}. This underscores that the warm gas kinematics is not the ideal observable to investigate the mechanisms driving turbulence.

\section{A predictive model for the evolution of turbulence} \label{sec:model}
\subsection{Feedback-driven model} \label{sec:sfrmodel}
In Sect.~\ref{sec:drivers}, we demonstrate that the $\sigma$-SFR distribution for galaxies in our extended sample aligns with the K18 feedback-driven models. These models, however, are based on assumptions about star-formation efficiency and the Toomre parameter (see Appendix \ref{sec:assumptionfeedback} for details). Additionally, the K18 models rely on a series of free parameters that are adjusted for different redshift ranges and specific galaxy populations. 
Given these complexities, in this section, we adopt the simpler feedback-driven model proposed by \citet[][hereafter F21]{Fraternali_2021}. The F21 model has fewer free parameters, enabling us to better control and understand their impact on the results. Despite its simplicity, the F21 model effectively describes galaxies at both low\citep{Bacchini_2020} and high redshift \citep{Rizzo_2021, Roman_2024}. This approach allows us to maintain accuracy while reducing the uncertainties inherent in more complex models. F21 assumes that supernovae must provide energy at a rate, $\dot{E}_{\mathrm{SN}}$, equal to the rate at which turbulent kinetic energy, $E_{\mathrm{kin}}$ is dissipated, that is, $\dot{E}_{\mathrm{SN}} = E_{\mathrm{kin}}/\tau$. The kinetic energy and supernova energy rate are defined as
      \begin{equation} \label{eq:ekin}
    E_{\mathrm{kin}} = \frac{3}{2} M_{\mathrm{gas}} \sigma^2
  \end{equation}
    \begin{equation} \label{eq:esn}
    \dot{E}_{\mathrm{SN}} = \epsilon_{\mathrm{SN}}\,\mathrm{SFR}\,\eta_{\mathrm{SN}}\,E_{\mathrm{SN}}\,
  \end{equation}
    where $\epsilon_{\mathrm{SN}}$ is the efficiency at which the kinetic energy is transferred to the ISM, $\eta_{\mathrm{SN}}$ is the supernova rate, $E_{\mathrm{SN}}$ is the supernova energy equal to 10$^{51}$ erg. The timescale of turbulence dissipation $\tau$ can be defined as $\tau = 2 h/\sigma$ \citep{Maclow_1998} where $h$ is the scale height of the gas disk. For a gas disk in hydrostatic equilibrium, the scale height can be written, to first-order approximation, as:
\begin{equation}
    h(R) = \sigma(R)\left[\frac{\partial{\phi(R,0)}}{\partial Z}\right]^{-1/2}
\end{equation}
where $\phi(R,0)$ is the galactic gravitational potential in the plane of the disk ($Z = 0$). Using $\dot{E}_{\mathrm{SN}} = E_{\mathrm{kin}}/\tau$, we derive the relation between $\sigma$ and SFR:
\begin{equation}
    \sigma = \mathrm{SFR}^{1/3} \left( \frac{4 \, \epsilon_{\mathrm{SN}} \, \eta_{\mathrm{SN}} \, E_{\mathrm{SN}}\,h}{3 \, M_{\mathrm{gas}}} \right)^{1/3}.
    \label{eq:slope0.33}
\end{equation}
Establishing the relationship between the local values of the SFR and $\sigma$ using Eq.~(\ref{eq:slope0.33}) is non-trivial due to the dependence of $h(R)$
not only on $\sigma(R)$ but also on the SFR via $M_{\mathrm{gas}}$, which affects the gravitational potential \citep{Bacchini_2020b, Ostriker_2022}. As a consequence the ratio $h/M_{\mathrm{gas}}$ may depend on both $\sigma$ and SFR. To investigate this, we used the scale height profiles $h(R)$ for 17 galaxies (4 at low-$z$ and 13 at high-$z$) from our extended sample. These profiles are presented and described in \citet{Bacchini_2019} and \citet{Bacchini_2024}, and they are derived using a method which assumes vertical hydrostatic equilibrium and takes into account the full galactic potential, including contributions from stars, dark matter, and gas \citep{Bacchini_2019}. We averaged these profiles radially and analyzed them in relation to $\sigma$, SFR, and $\sigma \times {\mathrm{SFR}}$. 
Additionally, we examined the distribution of $\epsilon_{\mathrm{SN}}h/M_{\mathrm{gas}}$ as a function of $\sigma$, SFR, and $\sigma \times {\mathrm{SFR}}$. This analysis revealed that the dependencies of $h$ and $\epsilon_{\mathrm{SN}}h/M_{\mathrm{gas}}$ on $\sigma$ and SFR cancel out when considering the average values of $h(R)$ and $\sigma(R)$ along with the global SFRs and $M_{\mathrm{gas}}$. Moreover, the factor $CC = [\epsilon_{\mathrm{SN}} (h/\mathrm{pc})/(M_{\mathrm{gas}}/M_{\odot}]^{1/3}$  remains approximately constant over cosmic time. The individual parameters in $CC$ evolve in such a way that they cancel out any redshift dependence.
For low-$z$ galaxies, $CC \sim 10^{-3}$ considering: $h \sim 200$ pc \citep{Bacchini_2020b}, $\epsilon_{\mathrm{SN}} \sim 0.01$ when global values of $h$ and $\sigma$ are considered \citep[see discussion in][]{Bacchini_2020}, $M_{\mathrm{gas}} \sim 10^9 M_{\odot}$ \citep[e.g.,][]{Leroy_2021b}. For high-$z$ galaxies, $CC \sim 10^{-3}$, considering: $h \sim 200$ pc \citep{Bacchini_2024}, $\epsilon_{\mathrm{SN}} \sim 0.01 - 0.1$ \citep{Rizzo_2020, Rizzo_2021, Roman_2024}, $M_{\mathrm{gas}} \sim 10^{10} M_{\odot}$. Throughout the rest of the paper, we adopt $CC$ as a constant value, leading to $\sigma \propto \mathrm{SFR}^{1/3}$, which differs from the K18 feedback models, which predict that
 $\sigma \propto \mathrm{SFR}^{1/2}$ or $\sigma$ remains constant.

\subsection{Calibrating the model} \label{sec:calibration}
In this section, we calibrate the physically-motivated model described in Sect.~\ref{sec:sfrmodel} using our extended sample. The final aim is to provide a relationship that predicts
the average turbulence, derived from cold gas tracers, for galaxies with known global SFRs.
To quantify the relationship between $\sigma$ and SFR, we fit a linear model between $\log(\sigma)$ and $\log(\mathrm{SFR})$ (see Appendix \ref{sec:fit} for details, Fig.~\ref{fig:sigma_sfr}), and we fixed its slope at 0.33, the value expected in the case of F21 feedback-driven turbulence. The analysis yields the following median relation:
    
    \begin{equation} \label{eq:sigma-sfr}
       \log{\sigma} = 0.33 \times \log{\mathrm{SFR}} + (0.77\pm 0.03),
    \end{equation}
    
with an intrinsic scatter of 0.13 dex. In Eq.~(\ref{eq:sigma-sfr}), $\sigma$ is in units of km s$^{-1}$ and SFR is in units of $M_{\odot}$ yr$^{-1}$. We note that, if the slope is left free to vary, we find a value of 0.27$\pm$0.03, consistent with the fixed-slope case within the 2-$\sigma$ uncertainties and $\gtrsim 8$-$\sigma$ larger than the slope of $\gtrsim 0.5$ predicted by K18 models (see Appendix \ref{sec:assumptionfeedback} for details). Furthermore, the leave-one-out cross-validation and the widely applicable information criterion methods \citep{Vehtari_2015} indicate that the model with the fixed slope is favored over the one with the free value. This is another confirmation of the supernova-driven nature of the turbulence. Based on eq.~(\ref{eq:slope0.33}), the relative small scatter in this relation confirms that the factor $CC = [\epsilon_{\mathrm{SN}} (h/\mathrm{pc})/(M_{\mathrm{gas}}/M_{\odot}]^{1/3}$ has small variations that do not correlate with the galaxy SFR and the galaxy mass. In Appendix \ref{appendix:correlation}, we also perform a diagnostic check to ensure that the residuals of the linear relation (Eq.~\ref{eq:sigma-sfr}) do not correlate with $\log{\sigma}$ and $\log{\mathrm{SFR}}$. Additionally, we estimate the partial correlation between these two quantities to isolate their relationship against residuals.

\begin{figure}[th!]
    \begin{center}  \includegraphics[width=\columnwidth]{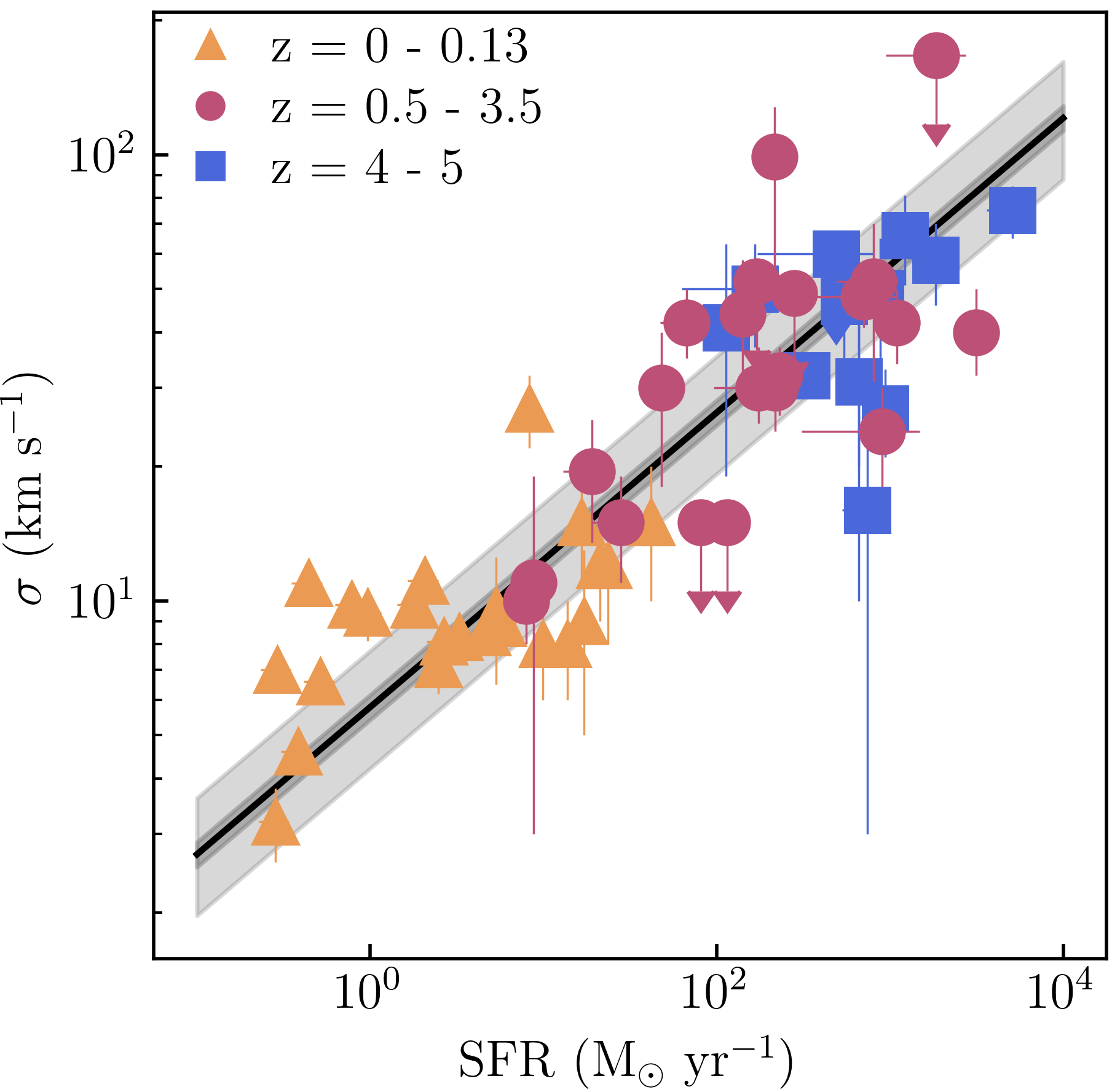}
        \caption{Distribution of $\sigma$ vs SFR for our extended sample divided in three redshift bins, as indicated in the legend. The black line shows the best-fit relation with the 1-$\sigma$ uncertainties (dark gray area) and the intrinsic scatter (gray area).}  
        \label{fig:sigma_sfr}
    \end{center}
\end{figure}

\subsection{Turbulence across cosmic time}
In this section, we present a semi-empirical model predicting the redshift evolution of turbulence when considering populations of main-sequence and starburst galaxies, at fixed $M_{\star}$ and $\Delta_{\mathrm{MS}}$. The main assumption of this model is that the evolution of $\sigma$ is mainly driven by the injection of energy due to the supernova explosions (see Sect.~\ref{sec:drivers}) and that, at fixed stellar mass, high-$z$ galaxies have higher SFRs than the local ones. By assuming the prescription of the redshift evolution of the SFR proposed by \citet[][Eq.~(\ref{eq:sfr})]{Popesso_2023}, and combining it with Eq.~(\ref{eq:sigma-sfr}), we derive the following evolution of $\sigma$ for galaxies at $z = 0 - 7$:
\begin{equation}
 \log{\sigma} (M_{\star}, z, \Delta_{\mathrm{MS}}) = 0.33 \times [\mathrm{MS}(M_{\star}, z) + \log{\Delta_{\mathrm{MS}}}] + 0.77.  \end{equation}

It is important to emphasize that this evolution of 
$\sigma$ as a function of redshift is entirely due to the evolution of the main-sequence relation. The physics of turbulence driving does not appear to evolve with time (see Fig.~\ref{fig:sigma_sfr}).

\begin{figure*}[th!]
    \begin{center}  \includegraphics[width=\textwidth]{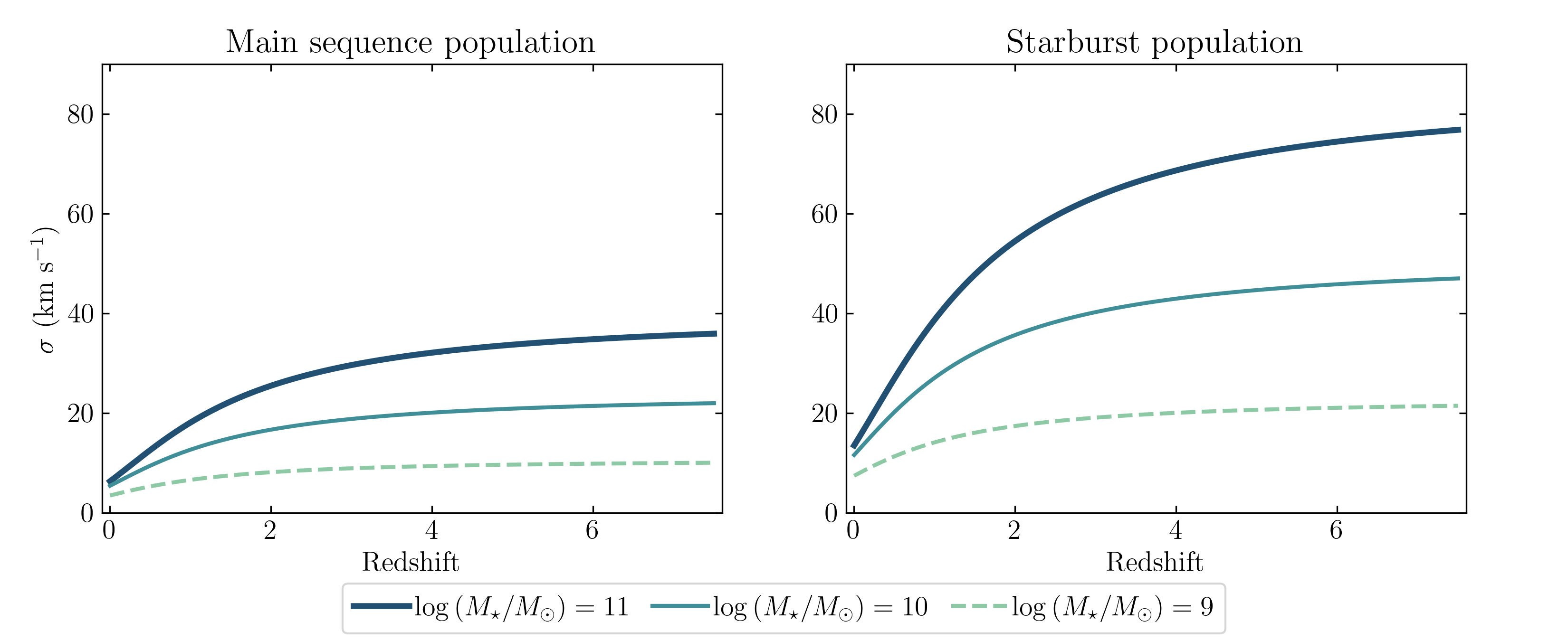}
        \caption{Predicted evolution of $\sigma$ as a function of redshift for populations of main-sequence ($\Delta_{\mathrm{MS}} = 1$, left panel) and starburst ($\Delta_{\mathrm{MS}} = 10$, right panel) galaxies. Different lines show the evolution at different stellar masses, as indicated in the legend.}  
        \label{fig:model}
    \end{center}
\end{figure*}

Fig.~\ref{fig:model} shows the redshift evolution for populations of main-sequence galaxies with $M_{\star}$ of $10^{11} M_{\odot}$ and  $10^{10} M_{\odot}$ in the left panel, while the right panel showcases the evolution for populations of starburst galaxies with $\mathrm{SFR}$ ten times higher than that of a main-sequence galaxy. We highlight the $\sigma$-$z$ relations in Fig.~\ref{fig:model} are not evolutionary tracks for individual galaxies, but profiles showing the evolution of $\sigma$ when galaxy populations with similar properties (i.e., $M_{\star}$, $\Delta_{\mathrm{MS}}$) are investigated. Interestingly, the supernova-driven model predicts that, at fixed redshift, main-sequence galaxies with lower stellar masses have lower velocity dispersions compared to galaxies with higher stellar masses, due to their lower SFRs.
We also show the evolution of $\sigma$ for galaxies with $M_{\star} = 10^{9} M_{\odot}$ with the dotted line. This model should be taken with caution as it relies on the extrapolation of the $\sigma$-SFR relation below the mass range probed by the extended sample. The values of $\sigma$ for the main-sequence galaxies at $z \gtrsim 4$ are consistent with the [CII]-derived values of $\sigma$ reported for galaxies from the zoom-in SERRA simulation \citep{Kohandel_2024}.

The models shown in Fig.~\ref{fig:model} suggest only a mild redshift evolution of turbulence motion within the range $z = 0$ - 2 and no significant evolution at $z > 2$. For instance, a main-sequence galaxy with $M_{\star} = 10^{10} M_{\odot}$ has $\sigma\approx$ 5 km/s at $z = 0$ to $\sigma\approx$ 16, 20, and 21 km/s at $z = 2$, 4, and 6. On the other hand, for a starburst galaxy with $\Delta_{\mathrm{MS}} = 10$, the model predicts $\sigma$ values of $\approx$ 10, 35, 46 km/s at $z = 0, 2$ and 6. For galaxies with smooth build-up history that grow in mass following the main-sequence relation \citep[e.g., the Milky Way; ][]{vandokkum_2013}, our model predicts that there is only a moderate evolution of turbulence over a large fraction of cosmic time. For instance, the Milky Way is expected to grow from $\approx 10^{10} M_{\odot}$
at $z = 2$ to a stellar mass of $\approx 5 \times 10^{10}$ at $z = 0$ \citep{Bland_2016}. During this time range, the average $\sigma$ values are expected to evolve from 16 to 6 km/s if the galaxy evolves following the main-sequence relation. Remarkably, the value of 6 km/s predicted by our model for a Milky-Way like galaxy at $z = 0$ is consistent with the average values of the velocity dispersion measured from molecular ($\approx$ 4 km/s) and atomic gas ($\approx 8$ km/s) \citep{Kramer_2016, Marasco_2017}.

\begin{figure}[th!]
    \begin{center}  \includegraphics[width=\columnwidth]{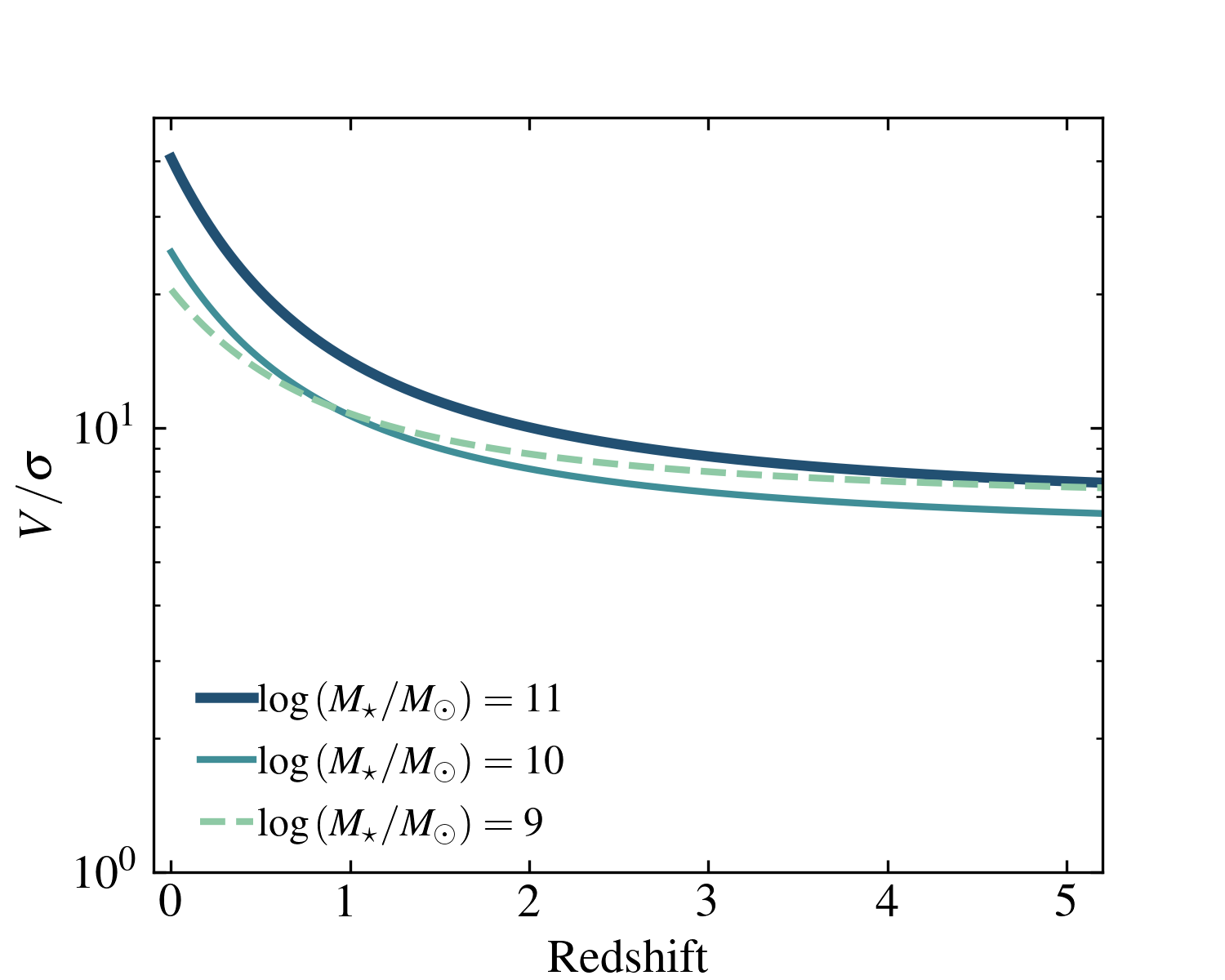}
        \caption{Evolution of the $V/\sigma$ ratio for populations of main-sequence galaxies. Note that at $z > 0$, these profiles should be considered as lower limits.}  
        \label{fig:tfr}
    \end{center}
\end{figure}

Predicting the evolution of the $V/\sigma$ across cosmic time is not straightforward, as it requires determining the evolution of the rotation velocity. The latter depends on the dynamical mass and the size of the galaxy. However, we can set lower limits for the evolution of $V/\sigma$ at fixed stellar mass by employing the local stellar Tully-Fisher relation, a scaling relation between the velocity and the stellar mass of a galaxy \citep{Tully, Reyes_2011}. The evolution of the stellar Tully-Fisher relation (TFR) has not been established within the redshift range of our extended sample; nonetheless, we can use the local stellar TFR to determine the lower limits of the velocity of galaxies at fixed stellar masses. This is because, within the $\Lambda$CMD cosmology framework, the stellar TFR is expected to evolve such that at fixed stellar mass, high-$z$ galaxies have higher velocity than $z = 0$ galaxies \citep{Dutton_2011}. 
In Fig.~\ref{fig:tfr}, we show the profiles obtained by dividing the velocity from the local stellar TFR from \citet{Reyes_2011} with respect to $\sigma$ from the supernova-driven model. This combination of the TFR and the supernova-driven model predicts that at $z > 1$, galaxies are rotationally supported with $V/\sigma$ of $\approx 10$, even for the low-mass ($M_{\star} = 10^9 M_{\odot}$) galaxy population.

\begin{figure*}[th!]
    \begin{center}  \includegraphics[width=\textwidth]{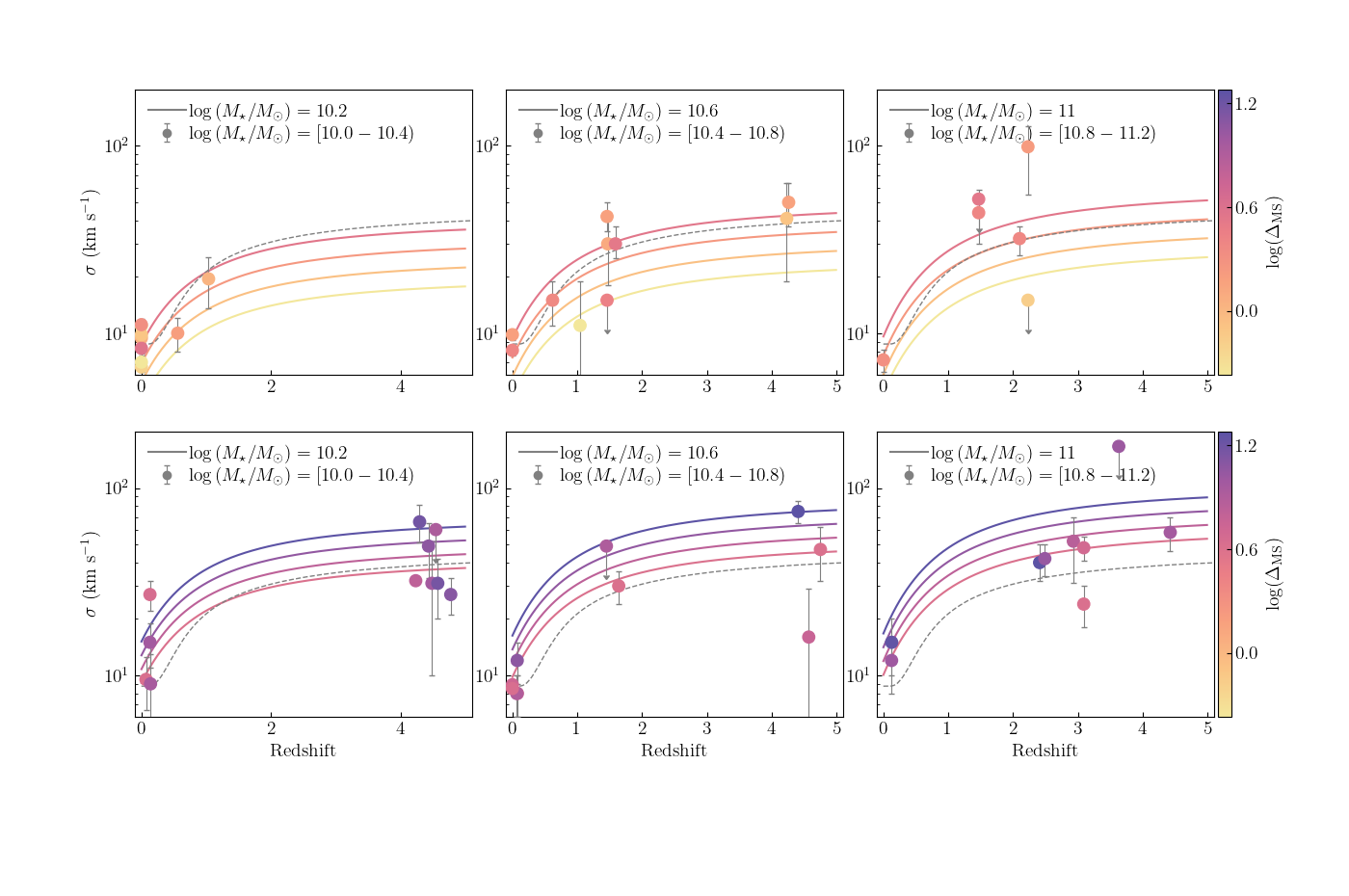}
        \caption{Comparison between the distribution of the extended sample in the $\sigma$-$z$ plane and the model predictions presented in Section~\ref{sec:model}. The upper and bottom panels show the main-sequence and starbursts galaxies, respectively. From left to right, the panels show observations and model curves in three stellar mass bins, as indicated in the legend. The data and the curves are color-coded according to the $\Delta_{\mathrm{MS}}$. For comparison, we show the best-fit empirical relation (see left panel in Fig.~\ref{fig:vs_cold}) with a gray dotted line. }  
        \label{fig:comparison}
    \end{center}
\end{figure*}

As a consistency check, in Fig.~\ref{fig:comparison}, we compare the distribution of the main-sequence (upper panels) and starburst (bottom panels) galaxies in the extended sample with the predictions from our model, by dividing the galaxies in three stellar mass bins. The model is in good agreement with observations, 
but the typical uncertainties on the $\sigma$ values and the low number of galaxies in each bin of stellar mass and $\Delta_{\mathrm{MS}}$ inhibit a detailed comparison. In the future, high-data quality observations (e.g., high S/N, angular resolutions) and larger sample sizes will facilitate the validation and refinement of this supernova-driven evolution model. For reference, in each panel of Fig.\ref{fig:comparison}, we show the empirical relation of $\sigma$-$z$ presented in Sect.\ref{sec:bestfit} with a gray dotted line. Its comparison with the supernova-driven models  highlights how this relation reflects the heterogeneity of the extended sample, which consists of galaxies spanning a wide range of $\Delta_{\mathrm{MS}}$.

\section{Caveats}\label{sec:caveats}
The results presented in this work rely on three observables: cold gas $\sigma$, warm gas $\sigma$, and SFR. In this section, we address potential caveats that may influence their measurements.

\subsection{Uncertainties on the velocity dispersion}
As discussed in Sect.~\ref{sec:sample}, deriving the kinematics of high-$z$ galaxies poses significant challenges. A primary limitation arises from angular resolution, often implying that the emission of high-$z$ galaxies is resolved by only few independent resolution elements. Consequently, the kinematics of high-$z$ galaxies can be influenced by the beam-smearing effect, leading to a degeneracy between $V$ and $\sigma$ and artificially inflated values of $\sigma$ 
 \citep{DiTeodoro_2015, Rizzo_2022}. For example, for about half of the extended sample at $z > 0.5$, the major axis is sampled by 3 - 4 independent resolution elements. While the kinematic analysis for all galaxies of our extended sample was performed using state-of-the-art forward-modeling techniques to mitigate the beam-smearing effect, overcoming this issue in the case of low-resolution data remains difficult. Consequently, velocity dispersion values may be slightly 
 overestimated. Moreover, the velocity dispersion at $z \gtrsim 1$ are $\approx$ 30 km/s, a value similar to the typical spectral resolution of the ALMA data. Thus, we cannot exclude that certain $\sigma$ values within our high-$z$ extended subsample may be slightly overestimated. This makes our conclusion even more robust. In fact, if the $\sigma$ values from cold gas were smaller, the discrepancy with the $\sigma$ values of warm gas would be even stronger.

The H$\alpha$ measurements used in Sect.~\ref{sec:comparison} are potentially affected by more severe  issues compared to the CO and [CII] data. Both in the case of low- and high-$z$ galaxies, the spectral resolutions of the instruments used for H$\alpha$ are worse than the ones used for cold gas observations (see also Sect~\ref{sec:intro}). For instance, the spectral resolution (FWHM) of the SAMI data is $\approx 70$ km/s, while the one for the KMOS data used by \citet{Birkin_2023} is $\approx 150$ km/s. In addition, the IFU data at $z \sim 2$ are taken in seeing-limited mode (resolution of $\approx$0.6", a factor of 3 worse than the ALMA data), resulting in H$\alpha$ emission which are only marginally resolved in some cases \citep{Hogan_2021, Birkin_2023}. As a result of these observational limitations, the warm gas $\sigma$ may be overestimated by some factor. This latter is not straightforward to quantify, as it would require the modelling of several mock datacubes at different spatial and spectral resolutions and for galaxies with different properties (e.g., size, inclination), a task beyond the scope of this work. 

\subsection{Different cold gas tracers}
In this study, we investigate the kinematics of cold gas (T$ < 10^4$ K) using a range of emission lines, including multiple CO transitions, [CI], and [CII]. While we expect minimal differences in the kinematics among various CO transitions and [CI] lines \citep[as discussed in][]{Rizzo_2023}, the [CII] emission line may exhibit variations compared to the kinematics traced by CO and [CI] lines. Having a lower ionization potential than HI (11.3 eV instead of 13.6 eV), [CII] can be generated in multiple phases of the ISM, including molecular, atomic, and ionized phases \citep[e.g.,][]{Rigopoulou_2014, Diaz_santos_2017, Wolfire_2022}, although several studies suggest that the majority (60 - 80\%) of [CII] emission arises from the molecular gas phase \citep[e.g.,][]{Pineda_2013, Vallini_2015, DeBreuck_2019}. As a rough estimate, if [CII] originates from warmer gas (T $\approx 10^2 -10^3$ K) than that traced by CO (T $\lesssim 10^2$ K), we estimate a thermal broadening speed of $\approx$ 2 km/s. Consequently, similar to CO at $z < 4$, the dominant contribution to the [CII] broadening ($\approx 30$ - 40 km/s) at $z = 4 - 5$ stems from gas turbulence. Conversely, if [CII] originates from the diffuse atomic and ionized phases, we expect [CII] velocity dispersion values that are, on average, larger than CO values. The absence of kinematic studies on both CO and [CII] for the same galaxies, both locally and at high-$z$, precludes us from providing quantitative estimates at present. However, future observations can potentially complement ALMA observations of [CII] at $z > 3$ with CO or [CI] observations using the high-frequency bands of the Square Kilometre Array \citep[SKA2,][]{Braun_2019} or the next generation Very Large Array \citep[ngVLA,][]{Kadler_2023}. Considering that the average [CII] $\sigma$ at $z > 4$ is only about 1.2 times larger than those measured at $z \sim 2$, any disparities between CO and [CII] kinematics would strengthen our findings, further attenuating the redshift evolution of turbulent motion.

\subsection{Considerations on SFR}
The model presented in Sect.~\ref{sec:model} relies on the $\sigma$-SFR relation. However, the SFR values of the extended sample are derived from diverse methodologies and assumptions as they are taken from the literature. We normalized the SFR values to a common IMF, but variations in assumptions and methodologies may introduce scatter \citep{Pacifici_2023}. 

Although radial profiles of $\sigma$ are available for the majority of galaxies within our extended sample, the absence of SFR surface density maps impedes an investigation of the efficiency of the turbulence drivers as a function of the galactocentric distance, as has been done for local galaxy samples \citep{Utomo_2019, Bacchini_2020}. Analyzing global $\sigma$ and SFR values enables an assessment of the primary mechanisms driving turbulence. Future high-quality observations mapping the H$\alpha$ or UV emission will allow us to obtain spatially-resolved maps of the SFR distribution and finally investigate the relative role of different driving mechanisms in the different regions of galaxies. 

\section{Summary and conclusions} \label{sec:conclusions}

In this study, we analyze the redshift evolution of galaxy dynamics (i.e., $\sigma$, $V/\sigma$) using kinematics from cold gas tracers (CO, [CI], [CII]). Our initial sample consists of 17 galaxy disks from the ALPAKA sample with high data quality. We complement ALPAKA with $\sigma$ and $V/\sigma$ measurements from the literature for subsamples of galaxies with spatially-resolved observations of emission lines tracing cold gas. This extended sample consists of 57 main-sequence and starburst disks at $z = 0 - 5$ and it covers a narrow range in stellar masses ($M_{\star} \approx 10^{10} - 10^{11} M_{\odot}$), but a wide range of 4 order of magnitude in SFRs. Our main results are the following.

\begin{itemize}

\item There is a trend of $\sigma$ increasing by around 2 to 3 times across the redshift range $z = 0 - 1$, while $V/\sigma$ decreases by a similar amount. However, these trends become less apparent at $z > 1$, where both $\sigma$ and $V/\sigma$ exhibit a plateau. We quantify the observed evolution using empirical relations, even though the derived trends reflect the heterogeneity of our extended sample, consisting of galaxies with a large range of SFRs.
\item We compare the kinematics of cold gas from our extended sample with the ones of warm gas. For the latter, we selected galaxies from the literature with physical properties (i.e., $M_{\star}$, SFR) matching the ones of our extended sample. The values of $\sigma$ ($V/\sigma$) from H$\alpha$ are a factor of $\approx 3$ higher (lower) than the ones from cold gas both for low and high-$z$ galaxies.
\item We investigate the potential drivers of turbulence and find that stellar feedback can reproduce the observed turbulence, traced by velocity dispersion measurements obtained from cold gas tracers. In contrast, gravitational instabilities tend to overpredict turbulence measured from cold gas. This finding diverges from previous studies based on warm gas kinematics, which suggested that turbulence is mainly driven by gravitational instabilities in $z > 1$ galaxies. 
\item We present a physically-motivated model predicting that turbulence is fed by the energy injected into the ISM by supernova explosions, and that the physics of turbulence driving does not to evolve with time. We calibrated this model on observations and developed a semi-empirical framework to predict the redshift evolution of turbulence in star-forming galaxies based on the evolution of the main-sequence relation. This model suggests a mild increase in turbulence within the redshift range 0 - 2 and no significant evolution at $z > 2$ for populations of galaxies at fixed stellar mass. Within this framework, for a Milky-Way-like progenitor, 
the velocity dispersion from cold gas evolve from $\approx 16$ km/s to 6 km/s in the last 10 Gyr ($z \lesssim 2$) of cosmic history. Interestingly, the supernova-driven model indicates that galaxies with lower stellar masses have lower velocity dispersion compared to more massive counterparts. By combining this model with the local stellar Tully-Fisher relation, we set lower limits for the redshift evolution of $V/\sigma$ at fixed stellar mass. We predict that high-$z$, low mass ($M_{\star} \gtrsim 10^9 M_{\odot}$) galaxy disks should be rotationally supported, with $V/\sigma \sim 10$.
\end{itemize}

The results presented in this paper indicate that the dynamical properties of galaxies significantly differ when derived from warm or cold gas.
This work highlights the importance of studying the evolution of galaxy dynamics, the turbulence within the ISM and its origin using cold gas tracers. 
The difference between cold and warm gas kinematics evidenced in our analysis may be due to (i) diffuse nature of warm, ionized gas compared to the cold one; (ii) bias of the velocity dispersion measurements to higher values due to contamination from non-circular motions due to
outflows; (iii) insufficient spectral and angular resolution. In the future, studies of the ionized gas kinematics using high angular and spectral resolution observations (FWHM $\lesssim 40$ km/s, e.g., VLT/ERIS; ELT/HARMONI) will be able to discriminate between these three hypotheses. 

The supernova-driven model presented in this paper makes testable predictions on the evolution of turbulence in galaxy disks across cosmic time. As more cold gas observations of star-forming galaxies with a wide range of stellar masses will be collected over the next years, we should be able to gain a more complete picture on the formation of the progenitors of disks in Milky-Way-like and local spiral galaxies.

\begin{acknowledgements}
The author thank the anonymous referee for constructive comments and suggestions. FR is grateful to Pavel Mancera Piña for useful comments and discussions.
FR acknowledges support from the European Union’s Horizon 2020 research and innovation program under the Marie Sklodowska-Curie grant agreement No. 847523 ‘INTERACTIONS’. This work has been supported by the Cosmic Dawn Center (DAWN), funded by the Danish National Research Foundation under grant No. 140. CB acknowledges support from the Carlsberg Foundation Fellowship Programme by Carlsbergfondet. 
LDM acknowledges support from the French government, through the UCA\textsuperscript{J.E.D.I.} Investments in the Future project managed by the National Research Agency (ANR) with the reference number ANR-15-IDEX-01. FRO acknowledges support from the Dutch Research Council (NWO) through the Klein-1 Grant code OCEN2.KLEIN.088. 
This paper makes use of the following ALMA data: 
ADS/JAO.ALMA\#2017.1.01659.S; ADS/JAO.ALMA\#2017.1.00471.S; ADS/JAO.ALMA\#2017.1.01674.S; ADS/JAO.ALMA\#2015.1.00862.S; ADS/JAO.ALMA\#2017.1.01228.S; ADS/JAO.ALMA\#2018.1.00974.S; ADS/JAO.ALMA\#2017.1.00413.S; ADS/JAO.ALMA\#2019.1.01362.S; ADS/JAO.ALMA\#2017.1.01045.S; ADS/JAO.ALMA\#2013.1.00059.S; ADS/JAO.ALMA\#2018.1.00543.S; ADS/JAO.ALMA\#2015.1.00723.S; ADS/JAO.ALMA\#2018.1.01146.S; ADS/JAO.ALMA\#2016.1.01155.S; ADS/JAO.ALMA\#2017.1.01677.S; ADS/JAO.ALMA\#2018.1.01306.S; ADS/JAO.ALMA\#2017.1.00908.S;
ADS/JAO.ALMA\#2012.1.00650.S;
ADS/JAO.ALMA\#2013.1.00803.S; ADS/JAO.ALMA\#2013.1.01161.S; 
ADS/ JAO.ALMA\#2015.1.00121.S; ADS/JAO.ALMA\#2015.1.00782.S; ADS/JAO.ALMA\#2015.1.00925.S; ADS/JAO.ALMA\#2015.1.00956.S; ADS/JAO.ALMA\#2016.1.00386.S; ADS/JAO.ALMA\#2017.1.00392.S; ADS/JAO.ALMA\#2017.1.00766.S; ADS/JAO.ALMA\#2017.1.00886.L; ADS/JAO.ALMA\#2018.1.00484.S; ADS/JAO.ALMA\#2018.1.01321.S; ADS/JAO.ALMA\#2018.1.01 651.S; ADS/JAO.ALMA\#2018.A.00062.S; ADS/JAO.ALMA\#2019.1.01235.S; ADS/JAO.ALMA\#2019.2.00129.S;
ADS/JAO.ALMA\#2016.1.01499.S;
ADS/JAO.ALMA\#2015.1.00330.S;
ADS/JAO.ALMA\#2017.1.00127.S; ADS/JAO.ALMA\#2017.1.00394.S;
ADS/JAO.ALMA\#2017.1.01052.S; ADS/JAO.ALMA\#2018.1.00001.S;
ADS/JAO.ALMA\#2017.1.01020.S; ADS/JAO.ALMA\#2019.1.00862.S;
ADS/JAO.ALMA\#2017.1.01471.S; ADS/JAO.ALMA\#2015.1.00456.S;
ADS/JAO.ALMA\#2011.0.00124.S;
ADS/JAO.ALMA\#2012.1.00978.S;
ADS/JAO.ALMA\#2015.1.01564.S.
ALMA is a partnership of ESO (representing its member states), NSF (USA) and NINS (Japan), together with NRC (Canada), NSC and ASIAA (Taiwan), and KASI (Republic of Korea), in cooperation with the Republic of Chile. The Joint ALMA Observatory is operated by ESO, AUI/NRAO and NAOJ.\\
We acknowledge usage of the Python programming language \citep{python3}, Astropy \citep{astropy}, Matplotlib \citep{matplotlib}, NumPy \citep{numpy}, and SciPy \citep{scipy}.

\end{acknowledgements}

\bibliographystyle{aa}
\bibliography{bib.bib}

\appendix

\section{Catalogue and table for the extended sample} \label{sec:details_sample}
With this paper we release\footnote{The information on the ALPAKA sources and the extended sample is available in electronic format at \href{https://alpaka-survey.github.io}{https://alpaka-survey.github.io}, where later releases and updates will also be placed.} a catalogue containing the physical properties for the 57 galaxies in the extended sample. 
The kinematic analysis of these galaxies is presented and discussed in previous papers (see Table \ref{tab:tab1}), with the exception of the PHANGS galaxies. For the latter, we used the combined 12-m and 7-m publicly available data cubes\footnote{\href{https://www.canfar.net/storage/list/phangs/RELEASES/PHANGS-ALMA/}{https://www.canfar.net/storage/list/phangs}} \citep{Leroy_2021b, Leroy_2021} to fit their kinematics using \bba. The parameter files used for the fitting and the corresponding outputs are available online \footnote{\href{10.5281/zenodo.10968307}{DOI:10.5281/zenodo.10968307}}.

\section{Fitting empirical relations} \label{sec:fit}
To fit Eqs.~(\ref{eq:sigma}), (\ref{eq:vsigma}) and (\ref{eq:sigma-sfr}), we use a hierarchical Bayesian approach\footnote{The analysis code --- \texttt{scattr} \citep{DiMascolo2024} --- is available at: \url{https://github.com/lucadimascolo/scattr}.} for the estimation of the likelihood function of the measured data \citep{Kelly_2007, Sereno_2016}. This hierarchical Bayesian method accounts for measurement uncertainties, intrinsic scatter, and upper limits. The sampling is performed by means of a Hamiltonian Monte Carlo (HMC; \citealt{Neal_2011}) inference approach, with No U-Turn Sampler \citep{Hoffman_2011} adaptation of the HMC step length.

\subsection{Fitting equations (\ref{eq:sigma}) and (\ref{eq:vsigma})}\label{sec:fit:1}
We aim at constraining the set of free parameters $\vec{\theta} = \{a, b, z_{0}\}$ of a given functional form $f(z;\vec{\theta})$ (i.e., Eq. (\ref{eq:sigma}) or (\ref{eq:vsigma})) described by a set of independent variables $z$ and observed quantities $y$ 
 (i.e., $\sigma$, $V/\sigma$). Here, we assume $z$ to represent a set of measurements not affected by intrinsic scatter and characterized by negligible uncertainties. This is related to the corresponding observed quantity $y$ through the relation

\begin{equation}
    y = \eta_y(z)+\epsilon_{y},
\end{equation}

where $\epsilon_y$ is the corresponding measurement uncertainty that follows a normal probability distribution with known variance $u^2_y$. The parameter $\eta_y$ instead represents the true value of the observed quantity $y$. To account for any potential scatter intrinsic to the measurement process or the measured quantities themselves, we assume $\eta$ to be drawn from the conditional distribution $p(\eta|z,\vec{\theta})$, assumed to be a normal density distribution $\mathcal{N}\{f(z;\vec{\theta}),s^2_{y}\}$ with mean equal to $f(z;\vec{\theta})$ and variance $s_{y}^2$. The latter term specifically quantifies the intrinsic scatter of the analysed data around the mean relation $f(z;\vec{\theta})$, and we include it as free parameter in the fit. The likelihood function of each measured data point is thus given by

\begin{equation}
    p(y|z,\vec{\theta}) = \int p(y|z, \eta) p (\eta| z,\vec{\theta}) d\eta.
\end{equation}
Since the data points are statistically independent, the full likelihood is the product of the likelihood function for the individual data points. For the prior probability of each model parameter, we assume wide uniform priors (see Table \ref{tab:priors}).

\begin{table}[!h] 
    \centering
    \caption{Ranges of the uniform priors used to perform the fitting of the $\sigma$-$z$ and $V/\sigma$-$z$ distributions using Eqs.~(\ref{eq:sigma}) and (\ref{eq:vsigma}).}
    \begin{tabular}{c|cc} 
    \hline
    \hline
    Parameter & $\sigma$ & $V/\sigma$\\
    \hline
       $a$  &    [0, 100] & [0, 100]\\
       $b$ &     [0, 100]  & [0, 100]\\
       $z_{0}$ &    [0, 10] & [0, 10]\\ 
       $s_{y}$ &   [0, 100] & [0, 100]\\
    \hline
    \end{tabular}
    \label{tab:priors}
\end{table}

\begin{figure}[th!]
    \begin{center}  \includegraphics[width=\columnwidth]{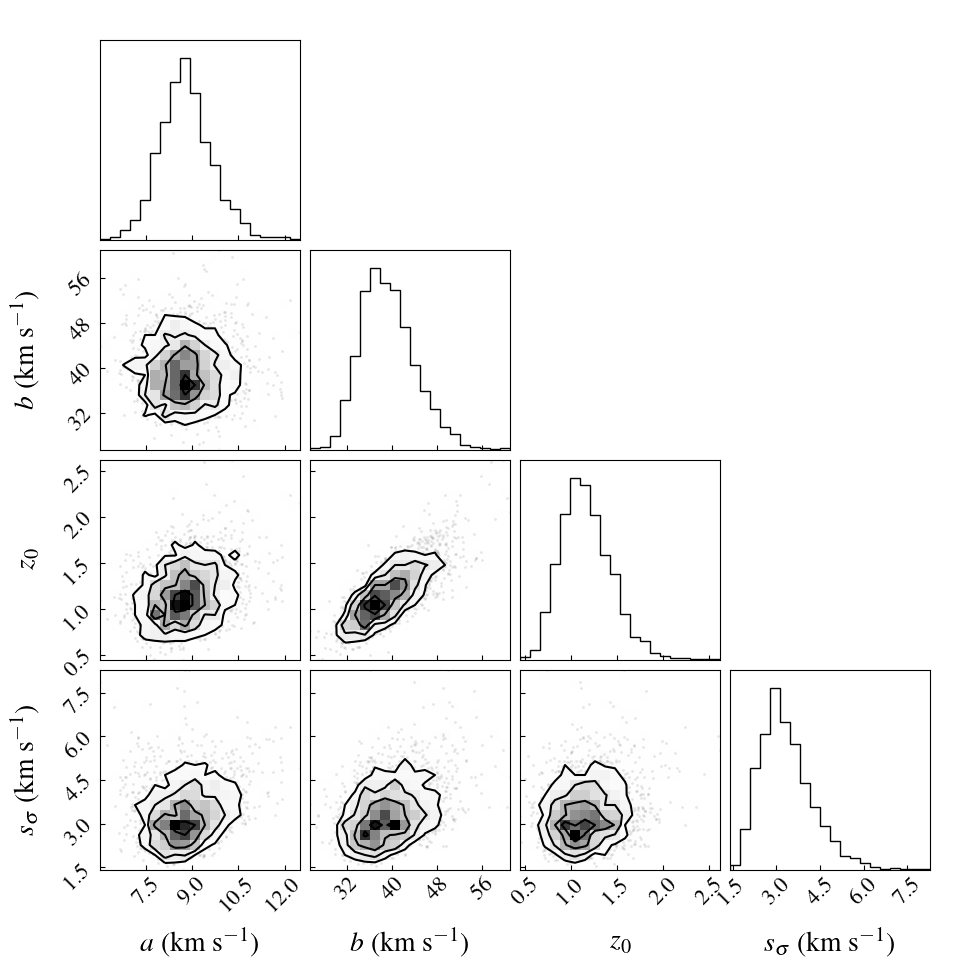}
        \caption{Posterior distributions for the parameter defining the empirical relation in Eq.~(\ref{eq:sigma}).
        }  
        \label{fig:posterior_sigma}
    \end{center}
\end{figure}

\begin{figure}[th!]
    \begin{center}  \includegraphics[width=\columnwidth]{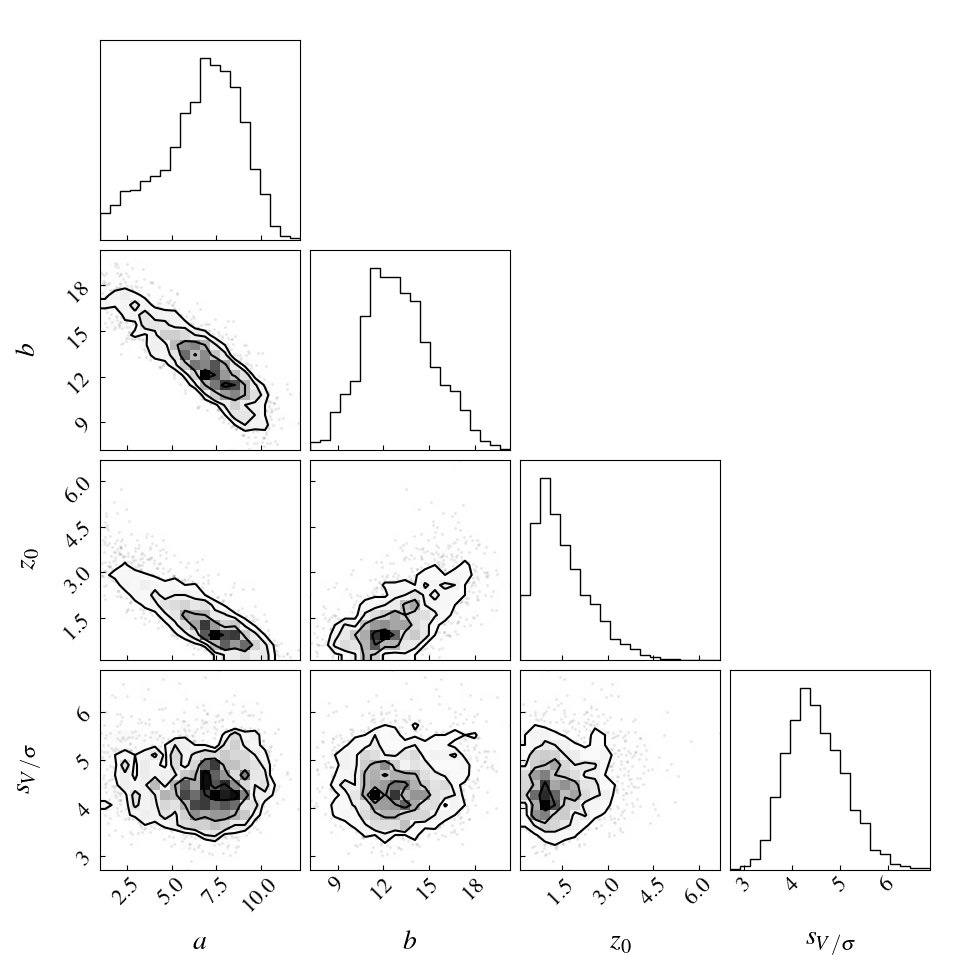}
        \caption{Posterior distributions for the parameter defining the empirical relation in Eq.~(\ref{eq:vsigma})}.
        \label{fig:posterior_vsigma}
    \end{center}
\end{figure}

\subsection{Fitting equation (\ref{eq:sigma-sfr})}\label{sec:fit:2}
In this case, both the independent ($\sigma$) and dependent variable (SFR) sets comprise observed quantities. As such, we should account for any associated measurement uncertainties and potential intrinsic scatter in both cases. To do so, we extend the hierarchical approach presented in the previous section to both the dependent and independent variables $x \equiv \sigma$ and $y \equiv$ SFR:
\begin{equation}
    \begin{split}
        x &= \eta_{x}(\xi)+\epsilon_x, \hspace{1cm} \eta_{x}=\mathcal{N}\{f_x(\xi;\vec{\theta}_x),s^{2}_{x}\}\\
        y &= \eta_{y}(\xi)+\epsilon_y, \hspace{1cm} \eta_{y}=\mathcal{N}\{f_y(\xi;\vec{\theta}_y),s^{2}_{y}\}.
    \end{split}
\end{equation}
In accordance with the notation in Sec.~\ref{sec:fit:1}, the variables $\eta_i$, $\epsilon_i$, denote the true values and measurement uncertainties of each variable $i=\{x,y\}$, respectively, and the corresponding functional model $f_i$, parameters $\vec{\theta}_i$ and intrinsic scatter$s^2_i$. Following \citet{Kelly_2007,Sereno_2016}, we further define $\xi$ as a reference independent variable drawn from a mixture of Gaussian probability distributions and related to the true values $\eta_{i}$ through the normal probability distributions $\mathcal{N}(f_i;s^2_i)$. In our specific case, 
\begin{equation} \label{eq:linear}
    \begin{split}
        f_x & = f_x(\xi) = 10^{\xi}, \\
        f_y & = f_y(\xi;b) = 10^{0.33\xi+b}.
    \end{split}
\end{equation}

For the prior probability on the offset parameter $b$ in Eq.~(\ref{eq:linear}), we assume a wide uniform prior in the range $[-100, 100]$. The joint likelihood function can instead be rewritten as

\begin{equation}
    p(x,y|\vec{\theta}) = \iint p(x,y|\eta_x,\eta_y) p(\eta_y|\xi,\vec{\theta}) p(\eta_x|\xi,\vec{\theta}) p(\xi) d\eta_x d\eta_y,
\end{equation}
where $p(\xi)$ is the underlying Gaussian mixture model. The term $p(x,y|\eta_x,\eta_y)$ represents the joint likelihood distribution for the $\{x,y\}$ variables at given $\{\eta_x,\eta_y\}$ and, such, could be interpreted as the probability function accounting for the uncertainties on each observed quantity. For most of the data in our sample, this can be described as a normal distribution with mean equal to the observed quantity $i$ and variance $u^2_i$ ($i=\{x,y\}$). However, for a subset of the analysed sample, we have access only to marginal constrain on the corresponding observed $y$ values. For these data points, we thus encode the available upper limits in the probability function $p(x,y|\eta_x,\eta_y)$ following the prescription by \citet{Sawicki_2012} and considering the cumulative distribution function of a normal probability distribution integrated up to the given upper limit. For a given set of $n$ independent data points $\{x,y\}$ with $m$ available upper limits on $y$, we can express the joint likelihood distribution as follows:
\begin{equation}
    \begin{split}
        p(x,y | \eta_x,\eta_y) 
             = &\prod^{n}_{k=1}\frac{1}{\sqrt{2\pi ~ u^2_{x,k}}}\exp{\left\{-\frac{1}{2}\left(\frac{x_{k}-\eta_{x,k}}{u_{x,k}}\right)^2\right\}} \\
        \times~ &\prod^{n-m}_{k=1}\frac{1}{\sqrt{2\pi ~ u^2_{y,k}}}\exp{\left\{-\frac{1}{2}\left(\frac{y_{k}-\eta_{y,k}}{u_{y,k}}\right)^2\right\}} \\
        \times~ &\prod^{n}_{k=m} \frac{1}{2}\left[1+\mathrm{erf}\left(\frac{y_{k}-\eta_{y,k}}{\sqrt{2}~u_{y,k}}\right)\right].
    \end{split}
\end{equation}

\section{Further comparison with previous works} \label{sec:ms}
\subsection{Warm gas kinematics from main-sequence galaxies} \label{sec:ms1}
In Sect.~\ref{sec:comparison}, we compare the redshift distribution of $\sigma$ from cold gas with the warm gas observations and models. Such comparison was conducted using observational samples that match the physical properties (i.e., stellar masses, SFR) of our extended sample. In this Section, we compare the cold gas $\sigma$ with the warm gas $\sigma$ derived using samples of main-sequence galaxies with a wider range of stellar masses than our extended sample.

\begin{figure}[th!]
    \begin{center}  \includegraphics[width=\columnwidth]{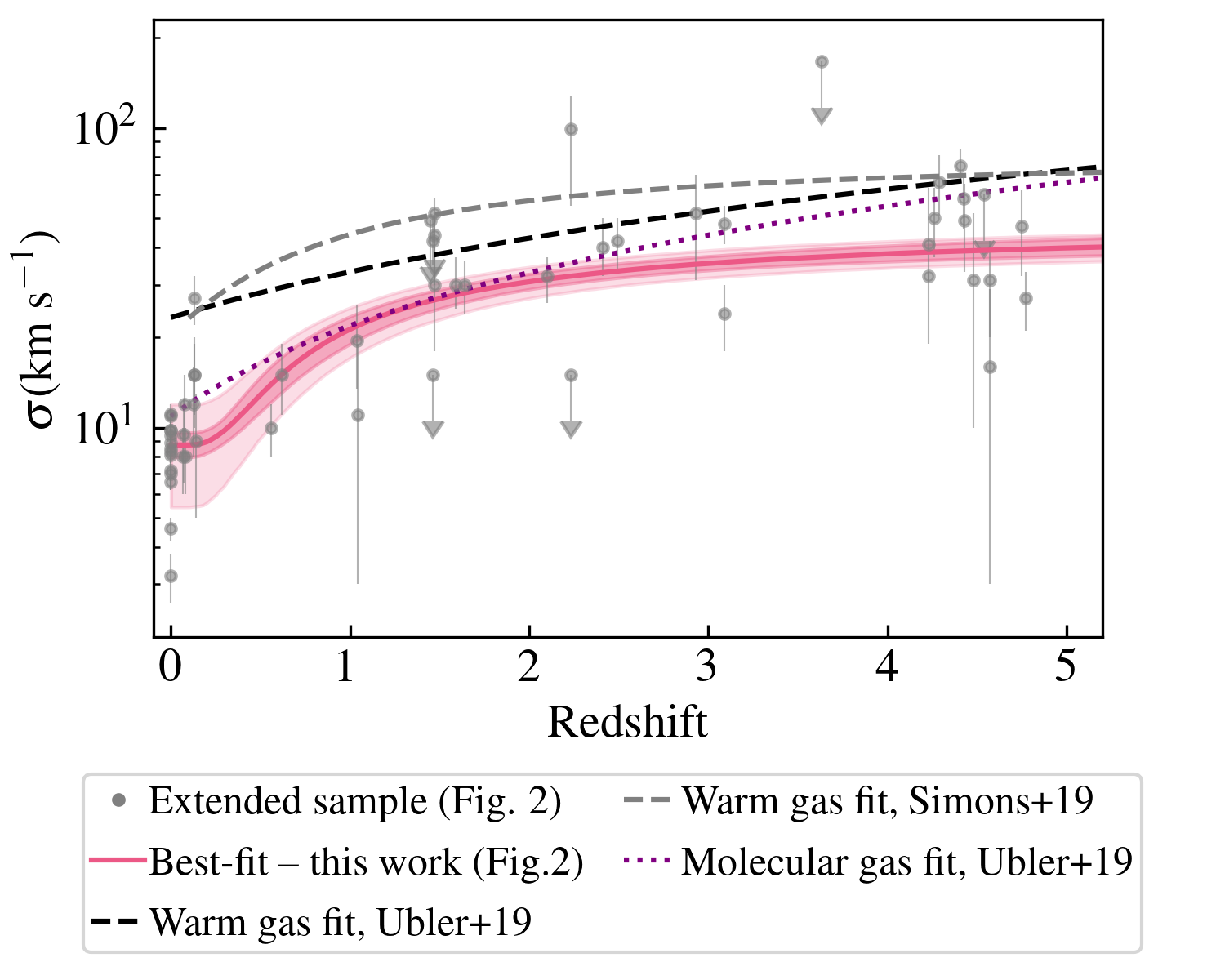}
        \caption{Comparison between cold gas dynamics and previous results from the literature. The gray points and pink lines show the extended sample and the best-fit relations presented in Fig.~\ref{fig:vs_cold}. The black and gray dashed lines show the best-fit relations found by \citet{Ubler_2019} and \citet{Simons_2017} using samples of main-sequence galaxies with stellar masses within a range larger ($\log(M_{\star}/M_{\odot}) = 9 - 11.5$) than the ones probed by the extended sample. The relations derived by \citet{Ubler_2019} and \citet{Simons_2017} are obtained by fitting data points in the redshift range [0, 3.8] and [0.1 - 2.5], respectively. The dashed curves show the extrapolations of these relations up to $z = 5$. Similarly, the purple dotted line shows the best-fit relation obtained by fitting molecular gas data for 9 main-sequence galaxies at $z = 0.7 - 2.4$ \citep{Ubler_2019} and its extrapolation up to $z = 5$.
        }  
        \label{fig:vs_main}
    \end{center}
\end{figure}

In Fig.~\ref{fig:vs_main}, the black dashed line represents the $\sigma$-redshift relation derived by \citet{Ubler_2019}. This relation was established by fitting warm gas $\sigma$ measurements from various IFU studies spanning $z = 0 - 3.8$ and comprising around 500 galaxies, while the line extension beyond $z = 3.8$ is based on an extrapolation. The compilation by \citet{Ubler_2019} includes main-sequence galaxies with stellar masses in the range $\log(M_{\star}/M_{\odot}) = 9 - 11.5$. 
The dashed gray line in Fig.~\ref{fig:vs_main} represents the $\sigma$-$z$ relation derived by \citet{Simons_2017} using H$\alpha$ slit observations from the DEEP2 \citep{Kassin_2012} and SIGMA \citep{Simons_2016} surveys, which include around 500 main-sequence galaxies with $\log(M_{\star}/M_{\odot}) = 9 - 11.5$ at $z = 0.1 - 2.5$. Interestingly, \citet{Simons_2017} find that the $\sigma$-redshift relation does not exhibit strong dependence on stellar mass, although simulations suggest that, on average, low-mass galaxies have lower velocity dispersions than high-mass galaxies \citep{Pillepich_2019, Kohandel_2024}. However, this discrepancy between observations and simulations might be explained by a systematic bias towards high $\sigma$ due to the beam-smearing effect (Sect.~\ref{sec:caveats}). This effect might be amplified in low-mass galaxies, which are typically only marginally resolved \citep[see discussion in][]{Ubler_2019}. Similarly to the results described in Sect.~\ref{sec:comparison}, both the relations by \citet{Ubler_2019} and \citet{Simons_2017}, derived from warm gas, systematically overestimate the values of the majority of CO/[CII] $\sigma$ of our extended sample. 

\subsection{Cold gas kinematics}  \label{sec:ms2}
To date, comprehensive investigations of the evolution of galaxy dynamics, leveraging kinematic data derived from cold gas tracers, remain largely unexplored due to the dearth of spatially-resolved observations for large samples of galaxies at redshifts beyond $z > 0$. An initial endeavor in this direction was undertaken by \citet{Ubler_2019}, who assembled a dataset comprising 9 main-sequence galaxies spanning the redshift range $z = 0.7 - 2.4$, with low-angular resolution ($\gtrsim 0.6$") CO observations \citep{Tacconi_2013, Ubler_2018}. The dotted purple line in Fig.~\ref{fig:vs_main} is from \citet{Ubler_2019} and represents the first attempt to establish an empirical relation describing the evolution of $\sigma$ with redshift using cold gas kinematics. The best-fit relation found by \citet{Ubler_2019}
is in overall agreement with the one obtained in Sect.~\ref{sec:bestfit} in the redshift range $z = 0.7 - 2$. However, \citet{Ubler_2019} do not provide an explanation for the data re-processing procedure used to infer the adopted CO $\sigma$ values, most of which are systematically lower ($\sim2\times$) than the original values reported in the first paper describing their kinematic analysis \citep{Tacconi_2013}.
In Fig.~\ref{fig:vs_main}, we show the extrapolation of the $\sigma$-$z$ at $z > 2.4$. As noted already by \citet{Rizzo_2021}, this extrapolation tends to overestimate, on average, the values of $\sigma$ measured at $z > 2$.

\subsection{Disk instability model and H$\alpha$ observations} \label{sec:comparison_di}
\begin{figure*}[h!]
    \begin{center}  \includegraphics[width=\textwidth]{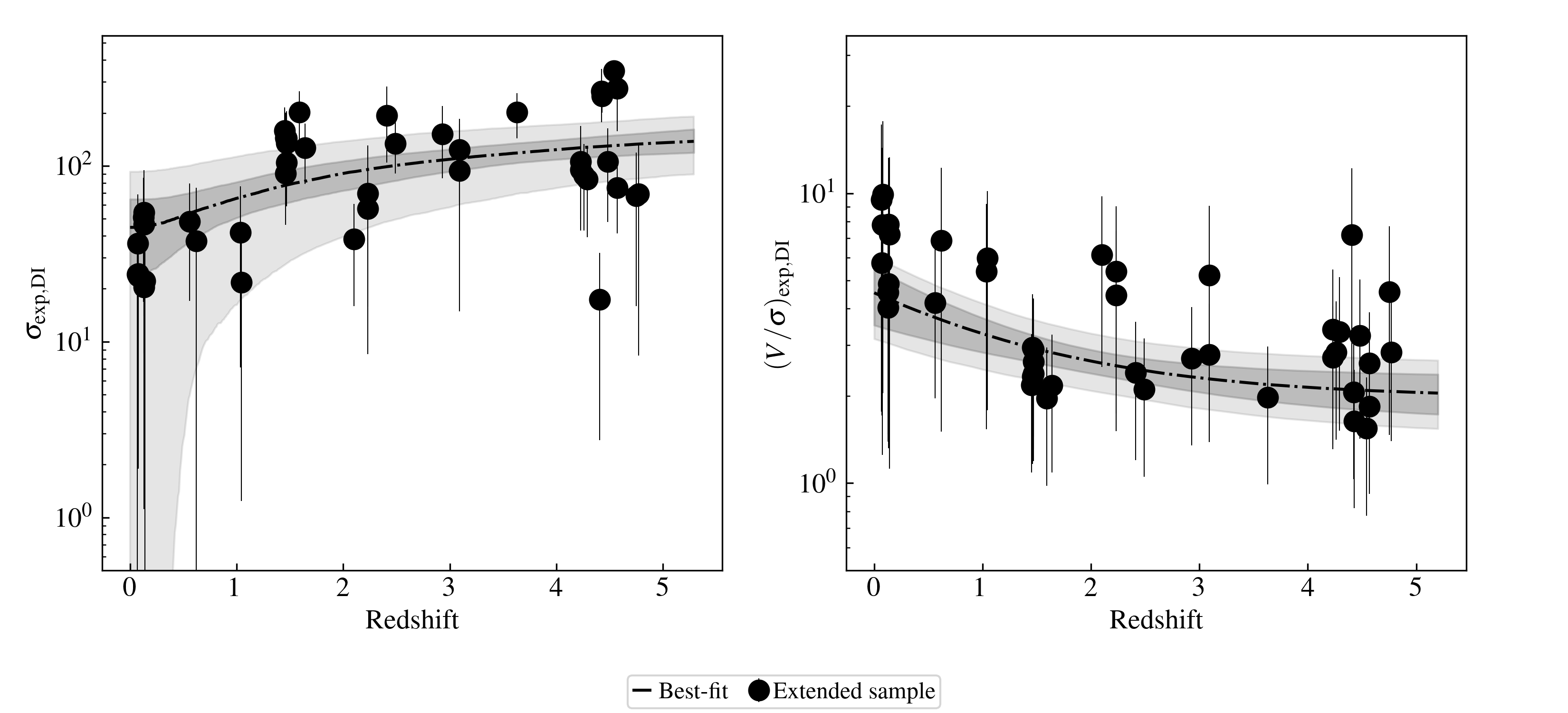}
        \caption{Redshift distribution of $\sigma_{\mathrm{exp, DI}}$ and $(V/\sigma)_{\mathrm{exp, DI}}$ for the low-$z$ starbursts and the $z \gtrsim 0.5$ galaxies in the extended sample. The black dash-dotted lines show the best-fit relations. }  
        \label{fig:fit_di}
    \end{center}
\end{figure*}

In Sect.~\ref{sec:comparison}, we compare the evolution of galaxy dynamics from cold gas with the predictions of the disk-instability model \citep{Wisnioski_2015}. To estimate the expected values of $\sigma_{\mathrm{exp, DI}}$ and $(V/\sigma)_{\mathrm{exp, DI}}$, we use Eq.~(\ref{eq:di1}) and Eq.~(\ref{eq:di2}) adopting our $f_{\mathrm{gas}}$  measurements. Given that $z = 0$ galaxies are not unstable (i.e., $Q \gg 1$), we compute $\sigma_{\mathrm{exp, DI}}$ and $(V/\sigma)_{\mathrm{exp, DI}}$ only for the galaxies that are expected to be unstable according to the Toomre instability criterion: low-$z$ starbursts \citep[e.g., see][for the analysis of $Q$ for the DYNAMO galaxies]{Girard_2021} and galaxies at $z > 0$ \citep{Wisnioski_2015, Turner_2017, Johnson_2018}. In Fig.~\ref{fig:fit_di}, we show the best-fit relations, obtained by fitting Eqs.~(\ref{eq:sigma}) and (\ref{eq:vsigma}) to the redshift distributions of $\sigma_{\mathrm{exp, DI}}$ and $(V/\sigma)_{\mathrm{exp, DI}}$.

\section{Assumptions of feedback models} \label{sec:assumptionfeedback}
\subsection{Krumholz et al. 2018}
The feedback models in K18 are derived under the assumption that the energy injected into the ISM by supernova explosions drives the turbulence. Unlike the approach taken by F21 (see Sect.~\ref{sec:sfrmodel}), K18 define the rate at which star formation adds energy to the gas not in terms of the energy injected by each supernova $E_{\mathrm{SN}}$ but in terms of the mean momentum injected per unit mass of stars formed ($p_{\star}/m_{\star})$, which is assumed to be 3000 km/s, based on numerical simulations \citep{Cioffi_1988, Thornton_1998, Martizzi_2015, Kim_2017}. The rate of energy injected by supernovae into the ISM per unit area is, therefore, defined, as
\begin{equation}
\dot{E}_{\mathrm{SN, K18}} = \Sigma_{\mathrm{SFR}} \left( \frac{p_{\star}}{m_{\star}} \right)\,\sigma,
\label{eq:snk18}
\end{equation}
where $\Sigma_{\mathrm{SFR}}$ is the SFR per unit area. Similarly to F21, the kinetic energy dissipated per unit area is defined as 
\begin{equation}
\dot{E}_{\mathrm{kin, K18}} = \frac{3}{2} \frac{\Sigma_{\mathrm{gas}} \sigma^3}{h}. 
\label{eq:kink18}
\end{equation}
In K18, Eq.~(\ref{eq:kink18}) is also expressed in terms of other model parameters, such as $Q$, $\Omega$ and the fractional contribution of gas to $Q$. A key difference between F21 and K18 is that, in K18, the rate of star formation per unit area is defined by the star formation law:

\begin{equation}
\Sigma_{\mathrm{SFR}} = \epsilon_{\mathrm{ff}} \frac{\Sigma_{\mathrm{gas}}}{t_{\mathrm{ff}}},
\label{eq:sfrk18}
\end{equation}
where $\epsilon_{\mathrm{ff}}$ is the star formation efficiency and $t_{\mathrm{ff}}$ is the free-fall time and $\Sigma_{\mathrm{gas}}$ is expressed as a function of $\sigma$ and $Q$, Eq.~(\ref{eq:toomre}).

In the K18 model, equations for $\epsilon_{\mathrm{ff}}$ and $Q$ are obtained by equating Eq.~(\ref{eq:snk18}) and (\ref{eq:kink18}). These two equations are then used to derive the relation between $\sigma$ and $\Sigma_{\mathrm{SFR}}$ and its integral SFR in two scenarios. In the first scenario, the disk is marginally unstable (i.e., value of $Q$ is fixed) and the star-formation efficiency is allowed to vary. By equating equations (\ref{eq:snk18}) and (\ref{eq:kink18}), it is found that $\epsilon_{\mathrm{ff}} \propto \sigma$, which, when substituted into Eq.~(\ref{eq:sfrk18}), results in $\sigma \propto \Sigma_{\mathrm{SFR}}^{1/2} \propto \mathrm{SFR}^{1/2}$. 
In the second scenario, the disk is assumed to be Toomre stable, with a fixed $\epsilon_{\mathrm{ff}}$ and variable $Q$. By equating Eqs.~(\ref{eq:snk18}) and (\ref{eq:kink18}), it is determined that $\sigma$ can only take one specific value, resulting in no direct relationship between $\sigma$ and $\mathrm{SFR} \propto \Sigma_{\mathrm{SFR}}$.

\subsection{Energy vs momentum}
In this section, we derive the relation between $\sigma$ and SFR by adopting the approach of F21, but using the momentum injected by supernovae rather than their energy. The energy injection rate is defined as

    \begin{equation}
\dot{E}_{\mathrm{SN, p}} = \mathrm{SFR} \left( \frac{p_{\star}}{m_{\star}} \right)\,\sigma .
\label{eq:snkf}
\end{equation}
By equating this expression with Eq.~(\ref{eq:ekin}), we obtain the following relationship:
\begin{equation}
\sigma^2 =  \mathrm{SFR} \left( \frac{p_{\star}}{m_{\star}} \right) \frac{4\, h}{3\, M_\mathrm{gas}}
\end{equation}
The primary difference from the F21 model lies in the slope. Here, $\sigma \propto \mathrm{SFR}^{1/2}$ when assuming a constant value for $\left(\frac{p_{\star}}{m_{\star}}\right)$ as in K18, or $\sigma \propto \mathrm{SFR}$ when assuming $p_{\star} \propto \sigma$. In contrast, F21 finds that $\sigma \propto \mathrm{SFR}^{1/3}$.

\begin{figure}[h!]
    \begin{center}  \includegraphics[width=\columnwidth]{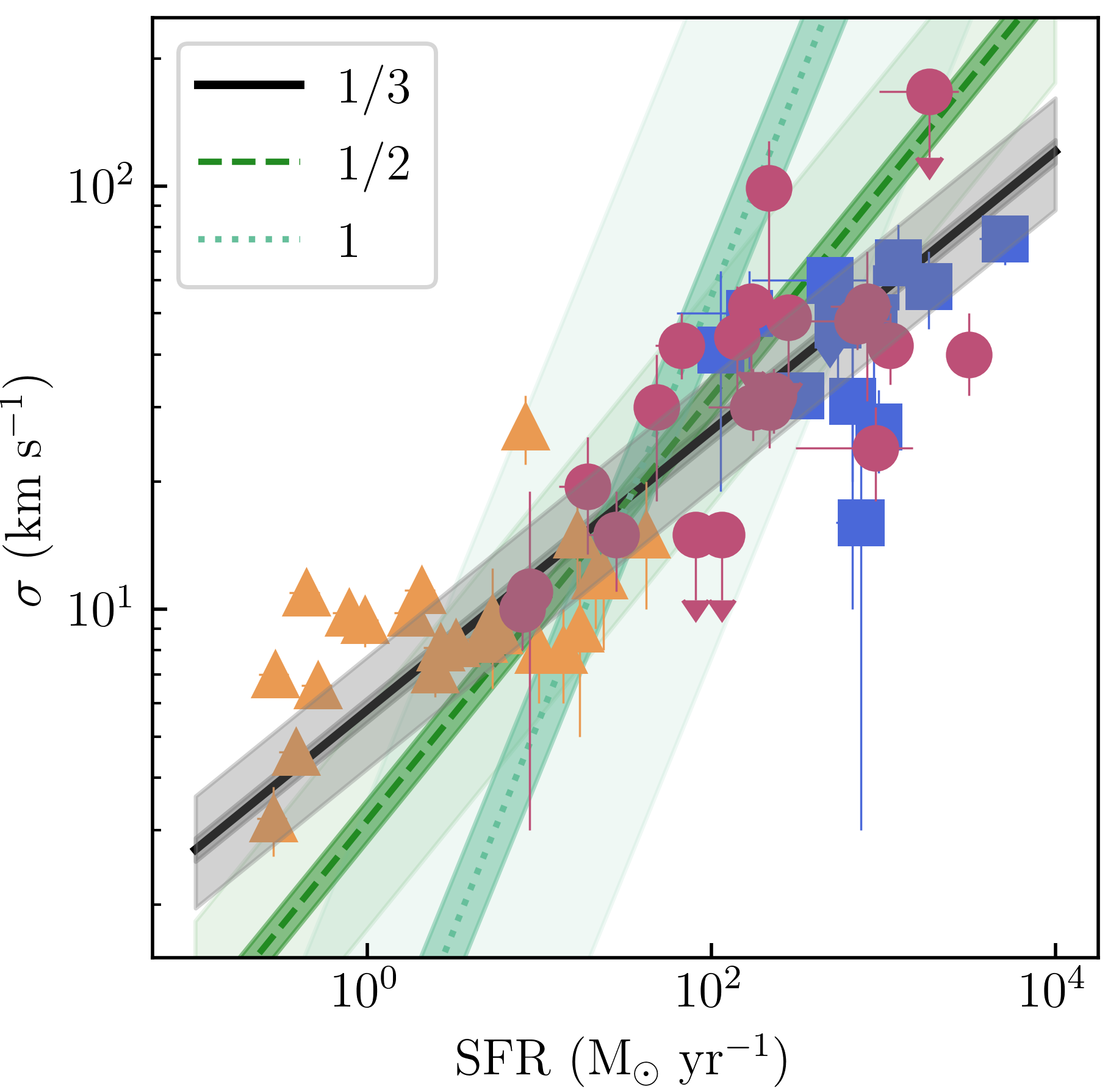}
        \caption{Same as Fig.~\ref{fig:sigma_sfr} but with the slope of the $\log{\sigma} - \log{\mathrm{SFR}}$ fixed at 0.5 (dashed, dark green line) and 1 (dotted green line).}
        \label{fig:slope}
    \end{center}
\end{figure}

For a consistency check, we fit the relation $\log{\sigma} = m \times \log{\mathrm{SFR}} + c$ with $m$ fixed at 0.5 (dark green dashed curve in Fig.\ref{fig:slope}) and 1 (light green dotted curve in Fig.\ref{fig:slope}). For comparison, we also include the fit with $m = 0.33$ (see Sect.~\ref{sec:model} and Fig.~\ref{fig:sigma_sfr}. As discussed in Sect.~\ref{sec:model}, the relation with a slope of $m = 0.33$ is the only one that accurately reproduces the data.

\section{Statistical analysis of the residuals}
\label{appendix:correlation}
\begin{figure*}[h!]
    \begin{center}  \includegraphics[width=\textwidth]{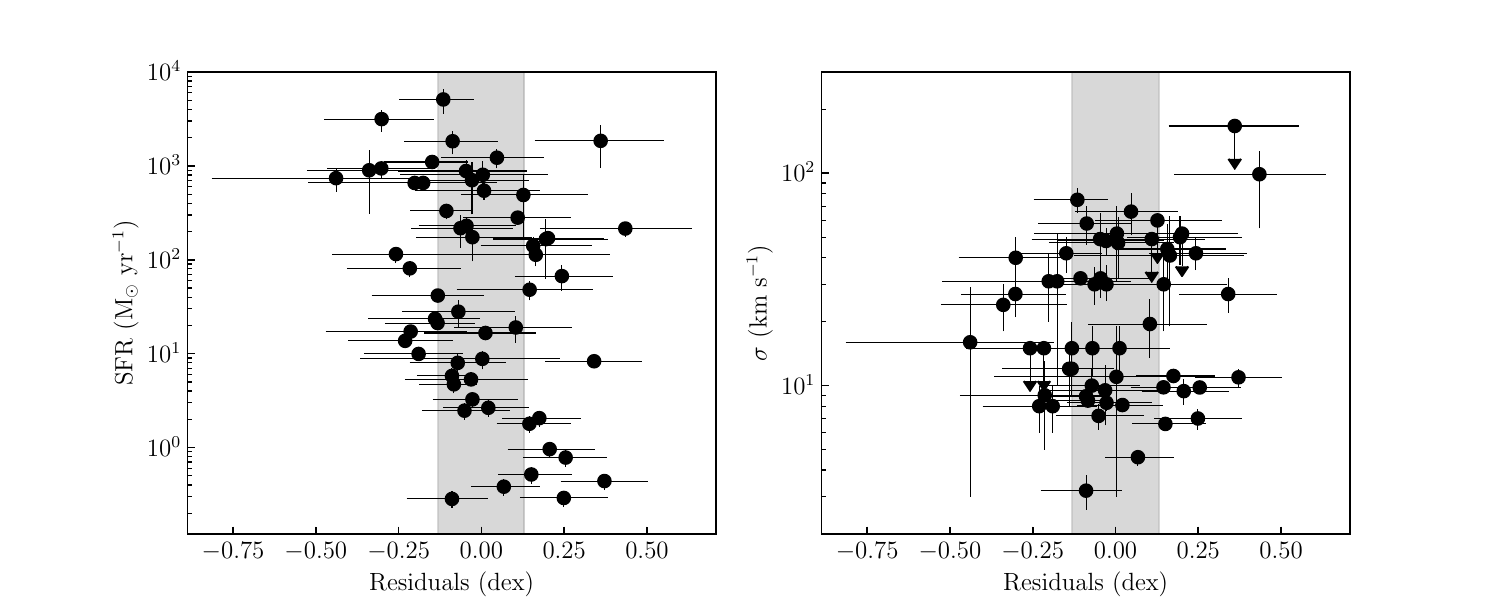}
        \caption{SFR and $\sigma$ as a function of the residuals from the best-fit relation defined by Eq.~(\ref{eq:sigma-sfr}). The gray shaded regions show the intrinsic scatter of the relation.}
        \label{fig:residuals}
    \end{center}
\end{figure*}

\begin{figure*}[h!]
    \begin{center}  \includegraphics[width=\textwidth]{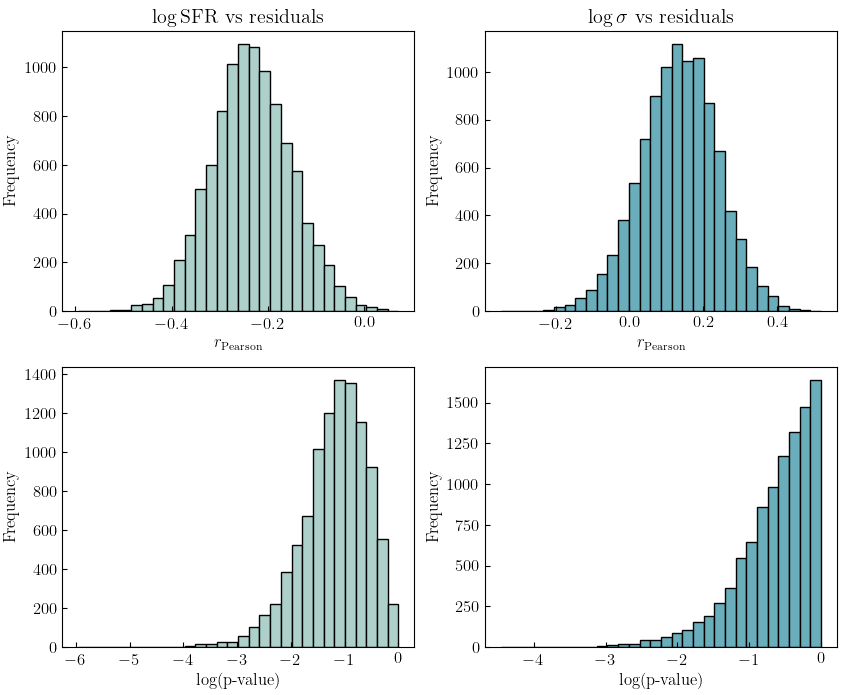}\caption{Distribution of the Pearson rank correlation coefficient (upper panels) and the p-value (bottom panels) between $\log{\mathrm{SFR}}$ and residuals (left panels) and $\log{\mathrm{\sigma}}$ and residuals (right panels). }  
        \label{fig:correlation}
    \end{center}
\end{figure*}

\begin{figure*}[h!]
    \begin{center}  \includegraphics[width=\textwidth]{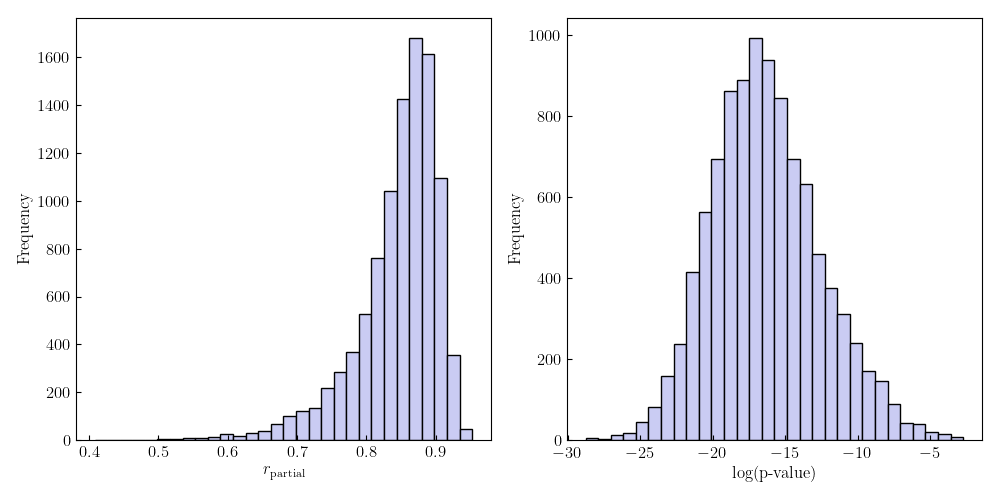}
    \caption{Distribution of the partial rank correlation coefficient (left panel) and the p-value (right panel) between $\log{\mathrm{SFR}}$ and $\log{\mathrm{\sigma}}$, while excluding the residuals. }  
        \label{fig:partial}
    \end{center}
\end{figure*}

In Sect.~\ref{sec:model}, we derived the expected relation between $\sigma$ and SFR (Eq.~(\ref{eq:slope0.33}) and discuss the potential dependence of $h$ and $M_{\mathrm{{gas}}}$ on $\sigma$ and SFR. As a diagnostic check, in this section we study the correlation between the residuals of the best-fit linear relations Eq.~(\ref{eq:sigma-sfr}) as a function of $\log\sigma$ and $\log\mathrm{SFR}$ (Fig.~\ref{fig:residuals}). Any correlations between the residuals and $\log\sigma$ and $\log\mathrm{SFR}$ would suggest that the linear model we are assuming with a slope fixed at 0.33 might not be appropriate. To perform this test, we compute the orthogonal distance between each data point ($\log\sigma_\mathrm{i},\log\mathrm{SFR_i}$) and the best-fit linear relation. To accurately propagate the uncertainties of the data points, we simulate multiple (N = 10000) datasets through bootstrapping. We then compute the distributions of the Pearson rank correlation coefficient $\rho$ \citep{Kendall_1973} between SFR and residuals and between $\sigma$ and the residuals, and their corresponding p-value distributions using a Monte
Carlo approach \citep{Curran_2014}. This methodology allows for robustly evaluating the correlation between quantities that have associated uncertainties. The resulting  distributions for $r_{\mathrm{Pearson}}$ and the p-values are plotted in Fig.~\ref{fig:correlation}. We find that the correlation coefficients are $r_{\mathrm{Pearson}} = -0.23_{-0.31}^{+0.15}$ for the $\log\mathrm{SFR}$ vs residual and $r_{\mathrm{Pearson}} = 0.13_{-0.03}^{+0.23}$ for the $\log\mathrm{\sigma}$ and residual distributions, respectively. The fractions of the two distributions with p-values $< 0.05$ are 36\% and 11\%, respectively. In other words, there is only a weak and no statistically significant correlation between residuals and $\log\mathrm{SFR}$, and between residuals and $\log\sigma$, respectively.  \\
As an additional consistency check, we calculate the partial correlation coefficients \citep{Kendall_1973} between $\log\mathrm{\sigma}$ and $\log\mathrm{SFR}$ while controlling for spurious or extra correlations beyond the linear relation by using the residuals as control variables. This analysis allows us to determine whether the correlation between $\log\mathrm{\sigma}$ and $\log\mathrm{SFR}$ is statistically significant or if it can be attributed to any correlations not accounted for in the adopted model. This is easily measured as a non-null correlation with the residuals, whose scatter and distribution could depend on $h/M_{\mathrm{gas}}$ and, therefore, on $\sigma$ and SFR. In Fig.~\ref{fig:partial}, we present the distributions of the partial correlation coefficients ($r_{\mathrm{partial}}$) and the corresponding p-values. The analysis yields $r_{\mathrm{partial}} = 0.86_{-0.06}^{+0.04}$. Based on the p-value distribution, where 100\% of p-values are less than 0.05, we conclude that the correlation between $\log\mathrm{\sigma}$ and $\log\mathrm{SFR}$ is statistically significant. This finding supports the robustness of the observed relationship, indicating that it persists even when accounting for the potential influence of residuals. Consequently, our results provide strong evidence that the relationship between $\log\mathrm{\sigma}$ and $\log\mathrm{SFR}$ is not merely an artifact of the residual effects of the assumed model.

\end{document}